*Executive Summary*

**February 27, 2018, Anthony W. Thomas (University of Adelaide)**

The IUPAP Working Group on International Cooperation in Nuclear Physics (WG.9) was established in 2003 at the IUPAP General Assembly in Cape Town, with a mandate to examine key issues in Nuclear Physics. Over the last 15 years the work of WG.9 has led to the creation of the Asian Nuclear Physics Association (ANPhA) and of the Association of Latin American Nuclear Physics and Applications (ALAFNA); members of WG.9 served on the OECD Global Science Forum Working Group on Nuclear Science which set out a roadmap for international nuclear science in 2008. The Working group also has close connections with NSAC in the United States and NuPECC in Europe and now organises regular meetings with funding agencies from around the world to promote nuclear science.

Through these activities the mandate of WG.9 has grown and its aims may now be summarized as:

- to provide a description of the landscape of key issues in Nuclear Physics research for the next 10 to 20 years;
- to produce (maintain) a compendium of facilities existing or under development worldwide;
- to establish a mapping of these facilities onto the scientific questions identified above;
- to identify missing components that would have to be developed to provide an optimized, comprehensive network of international facilities;
- to explore mechanisms and opportunities for enhancing international collaboration in nuclear science;
- to identify R/D projects that could benefit from international joint effort;
- to serve as a source of expert advice for governmental or inter- governmental organisations in connection with efforts to coordinate and promote nuclear science at the international level;
- to serve as a forum for the discussion of future directions of nuclear science in the broadest sense;
- to document the cross-disciplinary impact of Nuclear Physics and of nuclear facilities and to identify mechanisms for expanding (fostering) cross-disciplinary research.

The purpose of the present document follows closely that of the first IUPAP report on nuclear physics user facilities around the world, IUPAP Report 41. It has been significantly revised and updated and we expect that it will serve as a valuable resource for scientists and administrators.

In order to put the impressive array of nuclear physics facilities around the world into perspective, in this introduction we outline the key physics issues that have led to the facilities now operating, under construction or anticipated. In putting these physics issues forward we have drawn heavily on a number of recent reports, including the 2017 NuPECC and the 2015 NSAC Long Range Plans.

The NuPECC 2017 Long Range Plan provides a concise, accurate summary of the overall goal of nuclear physics, namely *"to unravel the fundamental properties of nuclei from their building blocks, protons and neutrons, and ultimately to determine the emergent complexity in the realm of the strong interaction from the* *underlying quark and gluon degrees of freedom of Quantum Chromodynamics (QCD)."* In order to do this one must explore more deeply the structure of hadrons, the forces between them, the limits of stability of nuclei and the properties of nuclear matter under extreme conditions. Advances in nuclear physics have important consequences for astrophysics, for physics beyond the Standard Model and for a rich bounty of applications in industry and health.

The following sections of this introduction provide background to the key questions which challenge modern nuclear physicists:

### 1) How do the structure of hadrons and their interactions emerge from QCD?

We already know a great deal about non-perturbative QCD but we are still far from a full understanding of the way quarks are confined inside hadrons. On the theoretical side lattice QCD has provided detailed information on the structure of the nucleon and its octet partners but the serious study of excited states is just beginning. We still do not know whether more than 3 quarks or one quark and one anti-quark can be confined or anything more complicated has a molecular nature. Both effective field theory and sophisticated model building have important roles to play in developing our understanding further. These theoretical ideas are being and will be tested and refined by an impressive array of measurements at new and upgraded facilities, most notably FAIR, JLab, J-PARC and RHIC. The relevant measurements vary from form factor and transition form factor measurements at ever increasing momentum transfer, to deep inelastic scattering (DIS) and semi-inclusive DIS, the latter made much more discriminating by clever new spin dependent measurements.

### 2) What is the structure of nuclear matter?

At the heart of this question is the structure and properties of atomic nuclei, the limits of existence as one moves from stability to proton or neutron drip lines, the changes in shell structure, the possible existence of an island of stability at masses above the heaviest nuclei currently known. We do not yet know what role if any the fundamental degrees of freedom of QCD, the quarks and gluons, play in finite nuclei, let alone in the far more dense matter that appears in neutron stars.

Amongst the latest facilities designed or built to tackle these issues experimentally we mention specifically EURISOL, FAIR, FRIB, ISAC I and II at TRIUMF and RAON under construction in South Korea. Theoretically there has been considerable progress in density functional theory, effective field theory, Green's function Monte Carlo methods and the no-core shell model.

### 3) What are the phases of nuclear matter?

Apart from finite nuclei, which form such an important part of modern nuclear physics, nuclear matter also appears in remarkably different conditions. In a neutron star one may reach densities up to six times the density of nuclear matter, where the composition is currently unknown; it could, for example, be quark matter with the boundaries between hadrons dissolved, or it could be nuclear matter with hyperons or some kind of condensed ate of kaons or superconducting quark matter. Some insight into this can be obtained in high energy heavy ion collisions, although in that case the matter is not in beta-equilibrium, as it must be in a star. In even higher energy relativistic heavy ion collisions one can form a high temperature quark-gluon plasma with remarkable properties, such a very low viscosity, which need to be further explored. Facilities such as the LHC at CERN (the ALICE experiment), RHIC at Brookhaven National Laboratory and FAIR in Germany are key to exploring this physics.

### 4) What is the role of nuclei in shaping the evolution of the Universe?

Most of the matter that we now take for granted, apart from deuterium, helium and lithium, was formed in stellar processes more than 3 minutes after the Big Bang. Some of these elements were formed as stars aged, others in supernova explosions and perhaps even in processes associated with neutron star mergers, which are so topical given the recent successes of the gravitational wave observatories. Understanding the origin and abundance of the elements is a fundamental objective of nuclear science. This is a field where there are wonderful opportunities for synergy with astronomy and astrophysics. However, the major new rare ion facilities currently operating, such as TRIUMF and RIKEN, as well as those under construction around the world, especially FAIR in Germany, FRIB in the US and RAON in Korea will be key to unravelling these mysteries.

### 5) What lies beyond the Standard Model

In spite of its manifold successes, the Standard Model suffers from many problems, not the least being a large number of ad hoc parameters, a huge fine-tuning problem and finally the fact that it describes only 5% of the stuff of the Universe. Nuclear techniques are vital to finding ways to detect dark matter and underground laboratories, where cosmic ray backgrounds are heavily suppressed, are crucial to the search for it. These laboratories are also vital to efforts to determine the nature of neutrinos, notably through searches for neutrinoless beta decay, as well as for numerous applications of nuclear science. Other topics actively being pursued include precision tests of the predictions of the Standard Model, notably involving parity violation and time reversal invariance.

# Nuclear Structure, Nuclear Reactions, and Nuclear Astrophysics

**Updated January 2018, by Alexandra Gade (Michigan State University)**

## 1. Introduction

Nuclei occupy the center of every atom, constituting 99.9 % of its mass. They are the building blocks of the visible matter in the Universe, of the Earth, and of us. The subfield of nuclear structure, reactions, and nuclear astrophysics strives to measure properties of nuclei, explain their existence, formation, and decays; use nuclei to explore physics beyond the Standard Model, and to meet societal needs in areas such as medicine, environment, and material science. The resulting broad research portfolio addresses the nature of the nuclear force that binds protons and neutrons into atomic nuclei as well as emerging dynamical many-body processes such as nuclear reactions or fission. The ultimate goal of the field is to develop a data-driven, predictive model of atomic nuclei and their interactions grounded in fundamental QCD and electroweak theory with quantified uncertainties. The ability to reliably calculate nuclear properties with the required precision will revolutionize our understanding of the origins of the chemical elements in the Universe and the complex reaction and decay networks that fuel explosive scenarios in the Cosmos. The importance of atomic nuclei to energy, health, environment, and security means that forefront research in this area plays a critical role in attracting and training the next generation of nuclear scientists needed in research, industry, and medicine.

The experimental work is carried out worldwide at major national (user) facilities and smaller, often University-based, laboratories employing a variety of accelerator technologies. While the large facilities continuously push the frontiers of the field, smaller laboratories are typically optimized for specific scientific programs, are engines of detector development and testing, and often allow for a unique hands-on education of students in all areas from acceleration to detection. The properties of rare isotopes and stable nuclei are interrogated with a powerful arsenal of experimental techniques, such as nuclear spectroscopy following nuclear reactions and decays, high-precision ion- and atom-trapping techniques, and laser spectroscopy, using state-of-the-art equipment ranging from hundreds-of-tons-heavy magnetic spectrographs and scintillator- and solid-state-based radiation detection arrays to table-top laser setups. Great strides have been made through nuclear spectroscopy in the continuum and for bound states. Penning-trap mass spectrometry has seen a leap in precision and laser-based techniques have broadened their realm of applicability. Experimental breakthroughs are often the result of new equipment developments or advances in accelerator technology, increasing the intensity and the reach for producing certain nuclei. New major rare-isotope facilities are under construction in the US and Europe, while existing, present-generation facilities continue to position the field for the next leap in discovery potential.

In recent years, nuclear theory has championed the development of nuclear many-body approaches whose underlying interactions are developed using chiral effective field theory with roots in the symmetries of QCD. Power counting and renormalization techniques have allowed inclusion of many-body correlations such as three-nucleon forces, which are not necessarily genuine but rather an inherent reflection of truncations in the order-by-order approach. Open-shell systems have now come into reach with such *ab-initio* type calculations for the first time. Developments in Lattice QCD started to connect to low-energy nuclear physics with the ultimate goal of anchoring the description of the very lightest nuclei and nucleon-nucleon interactions in QCD. Mean-field and configuration-interaction models have continued as pillars of the field in terms of describing experimental data and concluding on the driving forces of structural change in the exotic regime. The field of nuclear theory has moved to quantified uncertainties, including Bayesian approaches. This enabled meaningful comparisons of data and model calculations where uncertainties in assumptions and input parameters are encoded in theoretical error bars.

Weakly-bound nuclei are a fascinating open quantum system. The particle continuum is at the heart of rare-isotope science and methods to include the effects of weak or no binding near the nucleon drip lines continue being incorporated into the various theory frameworks. Computational physics, moving towards exa-scale computing power, has become the third leg of nuclear physics, dramatically and continuously widening the reach of nuclear many-body theory. The field of nuclear reactions has seen novel time-dependent approaches tackling "old" complex many-body processes such as fusion and fission. Reaction theory remains a field with crucial, open theoretical challenges that are a great opportunity for additional, new talent needed in this sub-field.

Due to the complexity of the underlying forces (QCD, QED, and electroweak) in nuclei, there have often been surprises. Progress in the field almost always has had experiment and theory working in close collaboration, informing each other about relevant nuclear properties to be measured or to be calculated for key nuclear systems. Properties of rare isotopes guide approximations, define parameters, and benchmark theoretical approaches. Discoveries continue to surprise, changing the paradigm of the field towards exciting new directions for experiment and theory.

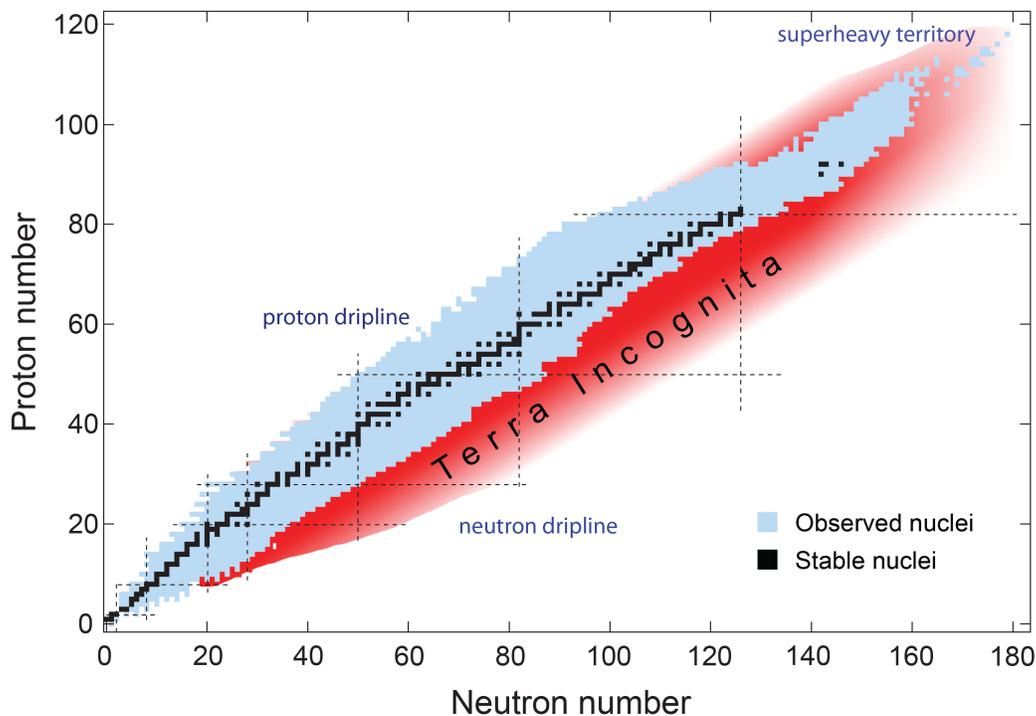

**Figure 1 - The nuclear territory** – Black: stable nuclei; blue: nuclei known to exist; red: Nuclei that will still be discovered. The chart of nuclei is bounded by the nucleon driplines where no additional nucleons can be bound. Superheavy nuclei reside at the limit of charge and mass. A vast, unknown, and still uncertain part of the nuclear chart, the terra incognita, remains unexplored.

## 2. The nuclear landscape

The territory of nuclear structure, nuclear reactions and nuclear astrophysics is the various forms of atomic nuclei that can exist. This landscape is illustrated by the chart of nuclei (Figure 1). Only 288 of several thousand nuclei, or isotopes, known to exist are either stable or practically stable (i.e., have half-lives longer than the expected life of the solar system). By moving away from the region of stable isotopes, by adding nucleons (either neutrons or protons), one enters the regime of short-lived rare isotopes, which are radioactive and decay by emitting β or α particles or disintegrate through the process of spontaneous fission. The limits of existence are demarcated by the nucleon driplines, beyond which additional nucleons cannot be bound anymore. The proton dripline has been reached for many isotopic chains. However, the neutron dripline is known only up to oxygen (Z=8). The superheavy nuclide with Z=118, A=294 marks the current upper limit of nuclear charge and mass. Today, about 3000 nuclides are known to us, but the number of those which have been well characterized is much less. At present, every year, the discovery of several 10 new isotopes is reported, near or beyond the proton dripline, in the neutron-rich regime, and in superheavy territory. Often, the decay properties of newly discovered isotopes can be studied, such as two-proton emission near the proton dripline or α decay chains for the superheavy elements produced for the first time. Beyond the neutron dripline, multi-neutron decays were studied as messengers of correlations inside such neutron-laden systems.

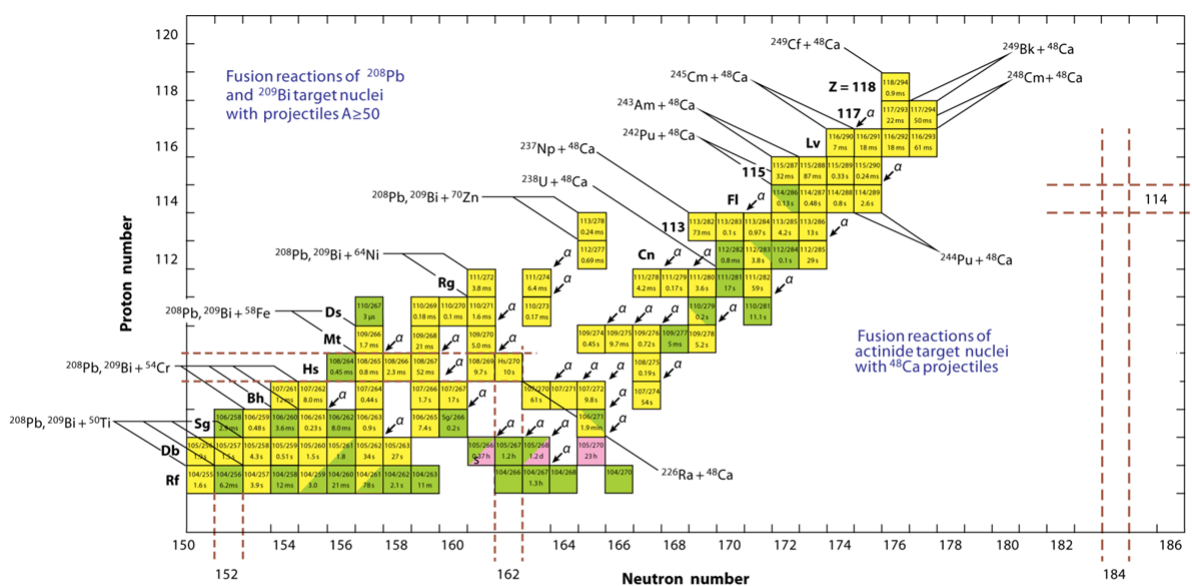

Hamilton JH, et al. 2013.
Annu. Rev. Nucl. Part. Sci. 63:383–405

**Figure 2 -** Superheavy territory, indicating some of the most recent discoveries, their production reaction, and decay chains.

The limits of atomic number continue to be expanded. Nine new elements have been discovered in the past 30 years. Recently, the International Union of Pure and Applied Chemistry IUPAC announced the names for four new elements. Nihonium (Nh), Moscovium (Mc), Tennessine (Ts), and Oganesson (Og), respectively, for elements 113, 115, 117, and 118. Enormous progress has been made in single-atom chemistry, characterizing for example Z=114 Flerovium and Z=112 Copernicium. As can be seen in Figure 2, most of the isotopes of elements of Z=113 and above were produced in the "hot" fusion of a Calcium-48 beam with actinide targets, the most exotic being Berkelium-249. At present, the pathway to reach beyond Z=118 is not clear. There may be more elements to be found once a means to produce them is realized.

Starting recently, atomic physics techniques have been used to study the heaviest elements: Two experiments have tried to observe X rays in coincidence with α decays to determine the atomic number, but were not successful. However, recently, laser spectroscopy of Nobelium succeeded, pointing to possible new directions. Based on the chemistry of heavy elements, the high atomic number of superheavy elements is expected to significantly alter their chemistry. There are speculations that beyond element 118 the electrons behave more like a Fermi gas and the chemistry for these heaviest elements will no longer follow a periodic table.

Exciting prospects exist for the expansion and study of the nuclear chart in all directions, with the Superheavy Element Factory under construction in Dubna and powerful next-generation rare-isotope beam facilities being completed in the US and Europe.

### 3. Nucleosynthesis in explosive scenarios – multimessenger revolution

Nuclear reactions and decays in stars, stellar explosions, and binary mergers generate energy and are responsible for the ongoing synthesis of the elements in the Universe. They are at the heart of many astrophysical phenomena, such as stars, novae, supernovae, nucleosynthesis in neutron-star collisions, and X-ray bursts. Nuclear physics critically determines the light curves of many objects, the signatures of isotopic and elemental abundances found in their spectra or in the composition of meteorites and presolar grains that originate from them, and the characteristic γ-ray radiation emitted by some of the objects. Neutron stars, the remnants of supernova explosions of massive stars, are among the strangest inhabitants of the Universe: ultra-compact, 1.5 times the mass of our Sun packed into an object with a diameter of merely 20 km, and with a crust that may be home to the most neutron-rich isotopes possible, surrounding an elusive core of unknown composition. The nuclear equation of state, electron capture rates, and the location of the neutron dripline are most important to understand these extreme objects. The field of nuclear astrophysics ties together nuclear and particle physics on the microscopic scale with the physics of stars and galaxies in a broad interdisciplinary context closely connected to astronomical observations and large-scale computation and theoretical modeling.

One of the grand science challenges of our time is the question for the origin of the heavy elements in the Universe. Given what we know about atomic nuclei, a rapid neutron capture process (r process) must be one of the major nucleosynthesis processes in nature. It is thought to produces roughly half of the nuclei found in nature beyond the Iron region. While many elements have contributions from multiple processes, there are some that are chiefly produced by an r process, such as Xenon, Gold, Platinum, and Uranium. The possible sites of the r process and consequently its conditions and reaction and decay sequences have remained elusive until 2017 where one r-process site was reported from a spectacular multimessenger astronomy campaign following the first LIGO/Virgo observation of the gravitational wave signal, GW170817, from a neutron-star merger. In addition to the gravitational-wave signal originating from two neutron stars spiraling into each other, a days-long glow, a kilonova, fueled by the radioactive decay of the synthesized neutron-rich nuclear matter, was left behind, observed and characterized across the electromagnetic spectrum. An understanding of the radioactive decays of the rare isotopes produced during the merger is crucial to interpret the kilonova signal and connect it to the observed cosmic abundances attributed to an r process. This observation has electrified the field of nuclear astrophysics and theoretical modeling of such events. Many β-decay measurements deep into the r-process relevant region have recently been performed at the Japanese rare-isotope leadership facility. The gravitational-wave signal from GW170817 has also provided unprecedented information on the size, mass, and deformability of neutron stars, which directly informs the fundamental properties of dense nuclear matter that can, to some extent, be probed in the laboratory through heavy-ion reactions during which nuclear matter is compressed to densities that depend on beam energy. Constraining the equation of state in laboratory experiments, for example, would allow to deduce the energy radiated by the gravitational waves. While it seems certain that these events produce lanthanides, the exact details of the relevant nuclear physics, most of which is yet unknown, will be needed to confirm that these sites are able to produce the heavier elements such as Gold and Uranium.

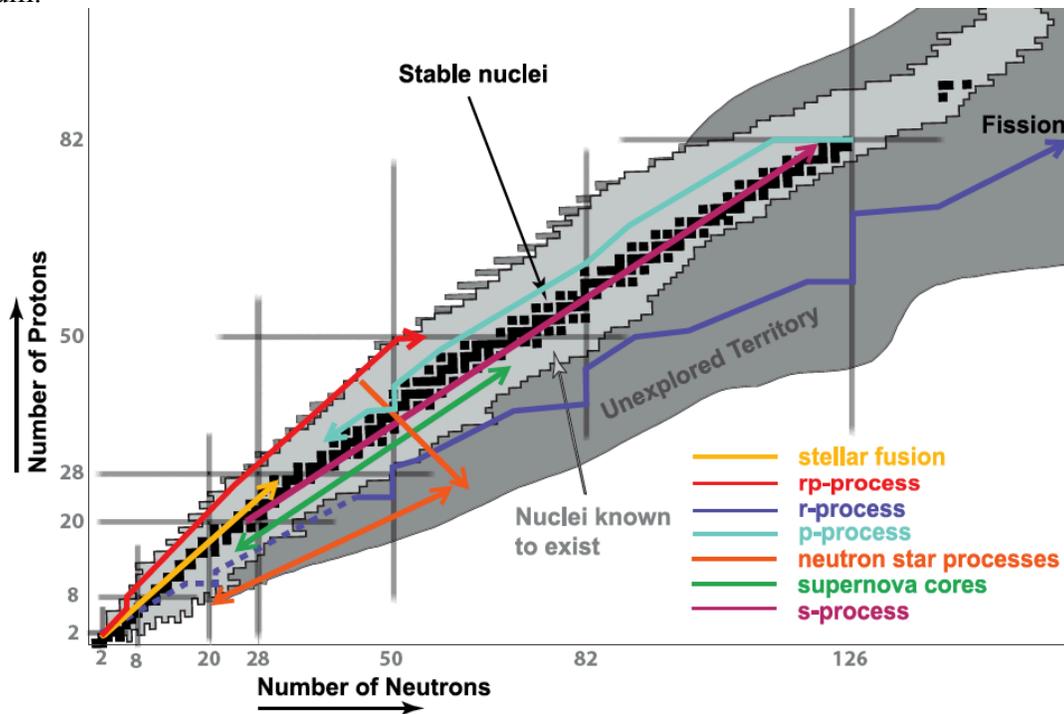

**Figure 3 -** Nucleosynthesis processes across the nuclear chart. For most, the reactions and decay properties of rare isotopes are critical to understand the abundance patterns left behind [Figure from F. Timmes].

Studies of other nucleosynthesis paths, such as the rapid proton capture (rp) process that fuels X-ray bursts on the surface of accreting neutron stars, are at the verge of a breakthrough with recoil separators existing or under construction at facilities that will be able to deliver high-intensity, low-energy rare-isotope beams to directly measure proton- and α-induced capture reactions along the rp-process path in the relevant energy regime. In the meantime, great strides continue to be made with indirect approaches, often based on transfer reactions that constrain important capture rates when direct measurements are still impossible.

## 4. New paradigms - not your textbook nuclear structure physics

At the heart of the physics of rare isotopes is the study of the particle continuum beyond the nucleon driplines or the nucleon separation energy, the interrogation of changes to the textbook nuclear structure that we know from detailed exploration of stable nuclei and the discovery, and study of new phenomena that are unique to extremely neutron- or proton-laden systems.

As the driplines or nucleon separation energies are approached, weak binding and the proximity of the continuum leads to unique correlations that probe nuclei in the context of open quantum many-body systems. Near the neutron continuum, in addition to studies of the neutron-rich Oxygen isotopes, possibly including the discovery of two-neutron radioactivity, the neutron spectroscopy of discrete unbound states was extended into the region of Neon and Magnesium, where shell evolution is at play and the heaviest neutron halo systems yet were uncovered. Neutron halos are a hallmark phenomenon of rare-isotope science and their recent identification in the A>30 region was a highlight. For heavier systems, one may not necessarily expect the formation of halos and it is rather suspected that the excess neutrons form a skin around a more symmetric core. While there is evidence for the development of neutron skins, the upcoming next-generation rare-isotope facilities are expected to reach nuclei with skins as thick as 0.5 fm, and after further upgrades even 0.8 fm. Reactions with such exotic systems may be the closest the field can get to study neutron matter in the laboratory.

One of the paradigms of nuclear structure is the shell model of the atomic nucleus, in which the motion of each neutron or proton is governed by a common force generated by all other nucleons. Nucleonic orbits bunch in energy, forming shells – and nuclei having filled nucleonic shells are exceptionally well bound. The numbers of nucleons needed to fill each successive shell are called magic numbers (2, 8, 20, 28, 50, 82, and 126 for stable nuclei). Among the most dramatic discoveries in current rare-isotope research has been the recognition that the traditional magic numbers can break down in the regime of extreme neutron-to-proton ratios and new shell gaps emerge. Recent progress has been made in the study of shell structure along the traditional magic chains Calcium, Nickel, and Tin and also in regions of rapid shell evolution such as neutron-rich Magnesium, Sulfur, Chromium, Iron, Selenium, and Krypton, for example. New detector arrays in the US and Europe as well has the highest beam intensities in Japan enabled the γ-ray and neutron spectroscopy of some of the most extreme systems of the isotopic chains mentioned above. Advances in mass measurements often allowed a first glimpse even before spectroscopic techniques were in reach, by looking at the evolution of nucleon separation energies, most importantly for the neutron-rich Calcium and Potassium isotopes, in recent years. The next-generation rare-isotope facilities under construction promise an unmatched reach for measurements of nuclear properties that indicate structural evolution and will allow to characterize in an unprecedented way the driving forces that need to be understood for a comprehensive model of atomic nuclei.

## 5. Rare isotopes as laboratory to unravel physics beyond the Standard Model

In spite of the stunning success of the Standard Model of particle physics, it is clear that there must be physics beyond. Collider experiments, at the LHC for example, aspire to probe extensions through the (anticipated) observation of new particles at the highest energy scale. Complementary approaches are afforded through the study of decays mediated by the weak force or through the observation of nuclear properties, such as an electric dipole moment, that break certain symmetries.

Neutrinos are fundamental within the Standard Model. They are the only fermions that do not carry electric or color charge. The only quantum number that can be used to distinguish between neutrino and anti-neutrino is the lepton number. There are many extensions to the Standard Model that do not require lepton number conservation. If the lepton number is violated, the distinction between the neutrino and the anti-neutrino becomes unclear, possibly making neutrinos Majorana fermions that are their own anti-particles. If the rare process of neutrino-less double β decay is observed, it is proven that the neutrino is of Majorana character. Preparations for ton-scale experiments are ongoing in the world, however, the nuclear matrix elements that govern the decay rate are highly uncertain at present. The nuclear matrix elements need to be known precisely for designing a neutrino-less double-β-decay search and – in the event of a positive observation – to deduce the neutrino's effective Majorana mass. The nuclear theory community has initiated concerted efforts to calculate the nuclear matrix elements with quantified uncertainties. Nuclear structure experiments – transfer and charge-exchange reactions – have been performed for the candidate nuclei, providing critical benchmarks for the nuclear structure models that aspire to calculate the neutrino-less double-β-decay matrix elements.

Addressing the apparent matter anti-matter asymmetry in the Universe, measurement of an electric dipole moment (EDM), which would separately violate parity and time-reversal symmetry, is one of the crucial probes of physics beyond the Standard Model, specifically searching for sources of CP violation. Heavy radioactive atoms hold promise as a sensitive system to search for EDMs. The EDM is induced by the interaction of the electrons with the nucleus. An enhancement of order 100-1000 (or more) is possible in nuclei that have pear-like shapes, such as Radium-225. Recently, experimental progress has been made in the trapping and manipulation of Radium-225 for an EDM measurement and the technical paths to new gains in sensitivity have been identified. One of the mid-term goals of the field is to identify and characterize the best candidates for sensitivity enhancements in atomic EDM searches. Next-generation rare-isotope facilities will provide candidate nuclei at much higher rate, promising orders-of-magnitude increase in sensitivity for EDM searches.

## 6. Relevance

The relevance of nuclear physics spans dimensions from $10^{-15}$ m (the proton radius) to objects as large as stars – it covers the evolutionary history of the Universe from fractions of a second after the Big Bang to today, 13.8 billion years later.

Nuclear physics, through uses of specific isotopes or derived technologies, impacts society in many ways. In addition to nuclear medicine, where certain radioactive isotopes are used therapeutically or for diagnostics, material and environmental science, energy research, or imaging techniques for security directly benefit from nuclear physics. Recent highlights include measurements of the decay heat of fission fragments of relevance for nuclear reactors. Basic nuclear physics techniques, such as β decay and total absorption measurements have made critical contributions. In the future, rare isotopes useful for a variety of application may be harvested at rare-isotope beam facilities in research quantities. Proof-of-principle experiments have already demonstrated enormous potential with great promise for the future.

Updated December 4, 2017, by Cédric Lorcé (Ecole polytechnique, Palaiseau)

This is a particularly exciting time for hadronic nuclear physics. While Quantum Chromodynamics (QCD), the fundamental theory of the strong interactions, has been thoroughly tested in the high energy or short distance regime, understanding the low energy or large distance regime remains one of the current key challenges. We are in the middle of a concerted effort to explore the consequences of non-perturbative QCD (the form of QCD applicable for strong binding), exploiting at the same time recent breakthroughs in lattice QCD, and new experimental facilities and machine upgrades to test those predictions. In this low energy regime, most of the attention has so far been drawn to the quark content, but a new major facility, the Electron-Ion Collider (EIC), is expected to be built in the next decade in the US to address for the first time in earnest the question of gluonic contributions, essential for completing and refining our modern picture of hadrons.

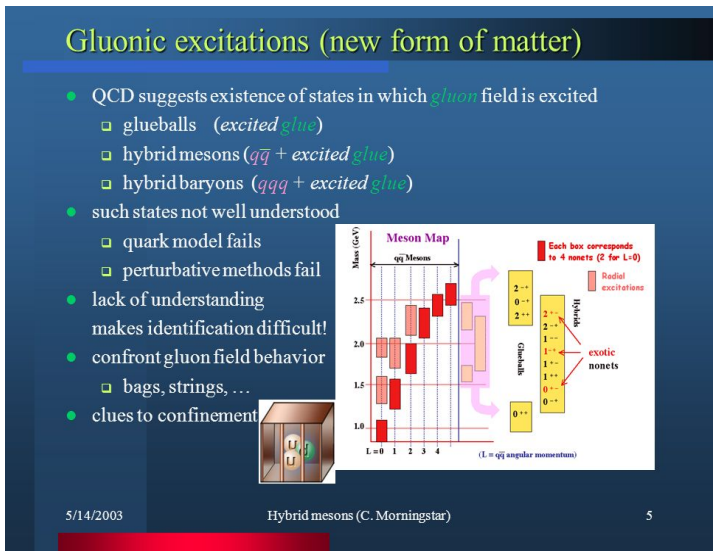

**Figure 1 – Spectrum of mesons (as anticipated by lattice calculations as well as QCD-inspired modeling) in the mass range of 1.5 to 2.7 GeV. Those nonets with $J^{PC}$ quantum numbers that cannot simply be $q\bar{q}$ systems (often called « exotic ») are labeled in red on the right.**

The first great challenge to be addressed is the question of confinement, that is, why free quarks have never been observed. This is connected in a fundamental way with the nature of the QCD vacuum, a complex medium of quark and gluon condensates and of non-trivial topological structure. A key to unraveling this mystery experimentally is the prediction that within QCD one should find so-called exotic mesons (called hybrids and glue balls), in which gluons play more than a binding role, by contributing to the observed quantum numbers in a characteristic way. The experimental search for these exotic mesons is focused on the GlueX experiment at Jefferson Lab, upgraded to 12 GeV in order to have the necessary energy reach, and in the future on the anti-proton storage ring at the new FAIR facility at GSI, which will allow the high resolution exploration of charmed exotic mesons. These two facilities complement each other in that the latter will explore the nature of confinement in a heavy quark system, where a model invoking a flux-tube picture is perhaps a reasonable starting point, while the former focuses on light, relativistic quarks, for which it is much more difficult to construct a simple physical picture at present. Beside the search for exotic mesons, it has also been proposed to look for the LHCb charmed 'pentaquark' using photo-production of J/ at threshold in Hall C at Jefferson Lab. This will shed a new light on the existence and properties of this non-standard type of matter, and in turn on the confining force.

Another key question in hadronic physics concerns the origin of the nucleon mass and spin. While the Higgs mechanism generates the mass of most of the elementary particles, the masses of light hadrons (which accounts for more than 99% of all matter around us) find essentially their origin in the strong interactions. The huge mass difference between bound states (like e.g. protons and neutrons) and their constituents (quarks and gluons) suggests that these systems are highly relativistic, and hence carry a significant amount of orbital angular momentum (OAM) already in the lowest state. It has now been fairly well established that about 25-30% of the nucleon spin is attributed to quark spin. Recently, a compelling evidence for a large and positive gluon spin contribution, though with sizeable uncertainty, has been found using new results from the Relativistic Heavy Ion Collider (RHIC) in the US. The quest is now focused on nailing down the gluon contribution with better precision and the orbital motion. This is a formidable task where semi-inclusive deep inelastic scattering (SIDIS) studies at Jefferson Lab and Drell-Yan process studied at RHIC and the COMPASS experiment at CERN are currently playing a major role. They are however not sufficient to resolve in particular the gluonic contributions, which is one of the main motivations for building an EIC.

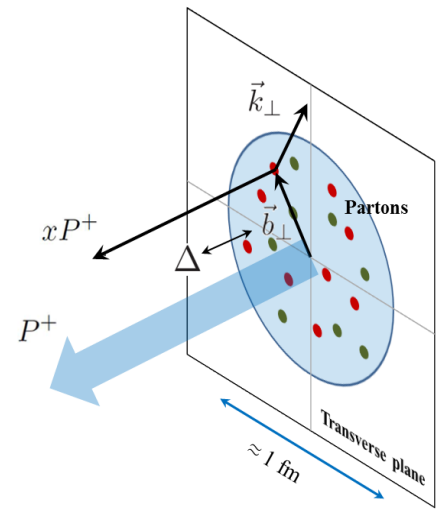

Figure 2 – Partonic picture of hadrons. Quarks and gluons are characterized by their momentum and/or position relative to their parent hadron

New experimental facilities offer new ways to explore familiar problems – sometimes with surprising results. The ability at Jefferson Lab to separate electric and magnetic form factors of the nucleon, by studying recoil polarization (rather than using the traditional Rosenbluth separation), has led to a dramatic change in the picture of the proton charge distribution. With the 12 GeV upgrade of CEBAF at Jefferson Lab, accurate and reliable measurements of both the electric and magnetic form factors of the proton and neutron can be extended to distance scales a factor of two smaller than currently possible. That is, measurements can be made of structures in the nucleon at distance scales much smaller than the nucleon itself, thereby probing the inner structure. By adding precision measurements of parity violation one can isolate the individual contributions of the u, d, and s quarks to these form factors – following pioneering work at MIT-Bates, MAMI@Mainz, and Jefferson Lab. Data on nucleon form factors in the time-like region are largely incomplete, but important results have recently been obtained by the BABAR collaboration.

In the last decade, there has been a huge interest in a new set of physical observables, the Generalized Parton Distributions (GPDs), which offer a three-dimensional (or tomographic) view of the internal structure of hadrons and, eventually, nuclei. Moreover, the GPDs also give access to the gravitational form factors which characterize the distribution of mass and pressure forces inside the nucleon. The CLAS 12 detector at Jefferson Lab has been designed to explore the proton GPDs across the entire valence region, while allowing sufficient overlap with the excellent work already done at smaller x at SLAC and reaffirmed in HERA and with Hermes at DESY. A particularly important milestone is the determination of the OAM carried by the u and d valence quarks. This is a key element of the question about how the proton spin is made up as discussed earlier. One may also hope to investigate the GPDs by applying them to the analysis of exclusive antiproton-proton annihilation into two photons at large energy and momentum transfer. It is proposed to measure crossed channel Compton scattering and the related exclusive annihilation process with various final states (scalar meson, vector meson, or lepton pair) at FAIR.

Another set of new physical observables, the Transverse-Momentum parton Distributions (TMDs), has also recently attracted a considerable amount of attention. They provide complementary three-dimensional pictures of the nucleon, and reveal the importance of initial and final-states interactions. In particular, the so-called T-odd TMDs, like the Sivers and Boer-Mulders functions, are predicted to change sign when extracted from a Drell-Yan process instead of SIDIS. This fundamental feature is currently being tested at Jefferson Lab and COMPASS. TMDs allow one to probe also the transversely polarized quark distributions and hence determine the tensor charge of the nucleon. This is an important quantity as it measures the extent of relativistic effects and enters the analyses of dark matter searches. TMDs have recently been generalized so as to include also the information encoded in the GPDs. In doing so, a direct link with the OAM has been obtained, which greatly helped settling a longstanding controversy about the form of the proper nucleon spin decomposition. It is not clear so far whether generalized TMDs can be extracted from scattering experiments, but several processes have been proposed and are currently under investigation.

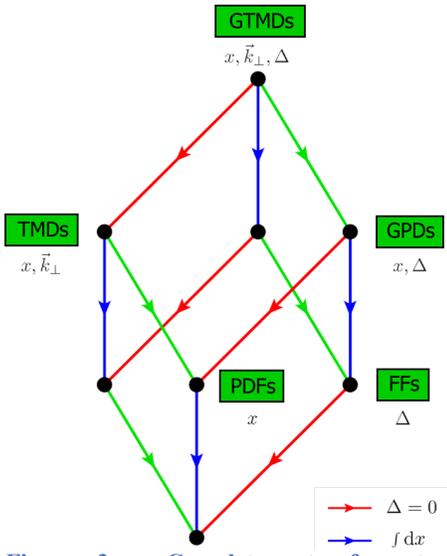

**Figure 3 – Complete set of parton distributions and their interrelations. $\Delta$ is the Fourier conjugate variable to position.**

A common theme across modern nuclear physics is the effect of a change of energy or baryon density on the QCD vacuum and the hadrons which may live in it. Studies at GSI have already yielded important information on the change of pion properties with baryon density. There are many theoretical predictions of changes of baryon and vector meson masses and other properties with density and temperature which must be tested experimentally. An array of techniques ranging from hadronic atom formation to antiproton annihilation in nuclei to the spin correlation parameters in quasi-elastic electron scattering will be applied to these issues over the next decade. Again the three new hadronic flag-ship facilities, FAIR, J-PARC, and Jefferson Lab with its 12 GeV upgrade, will carry the prime responsibility.

A particularly fundamental question of interest to both the nuclear structure and nuclear astrophysics communities are the origins of phenomena such as the observed 'saturation' of nuclear binding within QCD: that is the total binding energy of nuclei does not simply increase linearly with the number of nucleons, suggesting some kind of screening or reduction in nuclear interactions over the extended size of nuclei. Great progress has been made in understanding the stable nuclei in terms of effective two- and three-body forces, derived either phenomenologically or through modern effective field theory based on the symmetries of QCD. A truly microscopic understanding of the origins of these effective descriptions at the quark and gluon level would allow one gather more confidence in extending theories to regimes of density or neutron-proton asymmetry – for example at the densities found in the core of neutron stars or in nuclei with highly asymmetric numbers of neutrons and protons. Some progress has been made in relating the widely used Skyrme force to the quark gluon level description of nuclei and this work needs to be continued. However, most importantly, this consideration points to the experimental challenge of measuring the changes of the properties of hadrons immersed in a nuclear medium, discussed above, especially important.

Last but not least, tremendous progress has been achieved by the Lattice QCD community in the last few years. Calculations at the physical pion mass are now available, the so-called disconnected diagrams appear to give significant contributions to the flavor decomposition of angular momentum, and a new technique is currently under investigation allowing one to compute directly the momentum dependence of parton distributions. Effort will be invested in the near future to improve gluon contributions, and in particular the trace anomaly which plays a key role in the nucleon mass budget. No doubt that Lattice QCD will play an ever growing role in the study of the internal structure of hadrons.

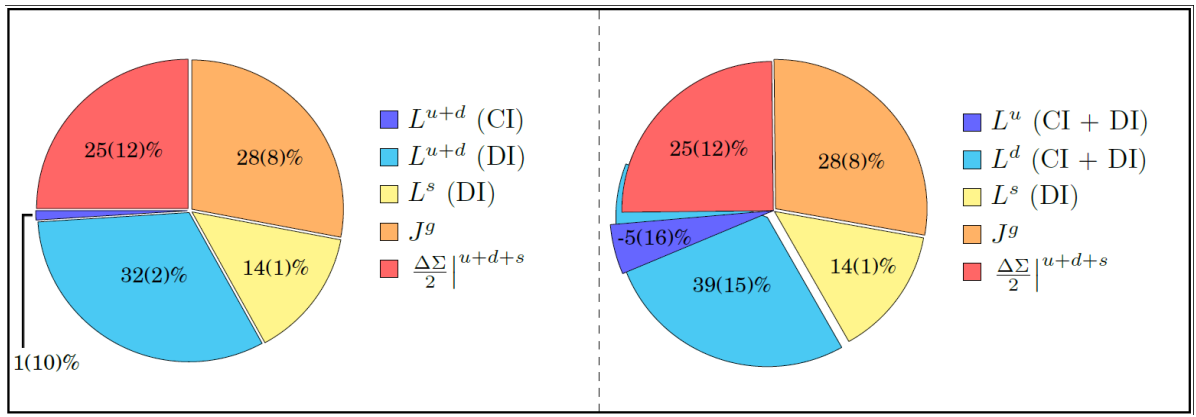

**Figure 4 – Pie charts for the quark and gluon spin and OAM fractions. The left panel shows the quark contributions separately for connected (CI) and disconnected insertions (DI), and the right panel shows the quark contributions for each flavor with CI and DI contributions summed together**

**Updated January 15, 2018, by Berndt Mueller (BNL)**

The fundamental interactions of nature are described by theoretical models called gauge theories. Quantum Chromodynamics (QCD), the theory of strong interactions, is such a theory, with just a small number of parameters to be determined from experiment. It is, however, difficult to compute the manifestations of QCD in nature except in the high momentum or short distance limit, in which the effective coupling is weak and where the perturbative approach is extremely successful. The spectrum and structure of strongly interacting particles is not calculable in this approach. Tools for understanding this structure include experiment and computer simulations of QCD. The important questions about QCD that can be accessed with these tools may be formulated as follows:

1. *What are the phases of strongly interacting matter and what roles do they play in the cosmos?*
2. *What is the role of gluons in nucleons and nuclei?*
3. *What determines the key features of QCD; can they be understood as manifestations of holographic duals described by gravity or string theory?*
4. *Can the study of QCD vacuum fluctuations at high temperature illuminate other early-universe processes, such as the one responsible for the matter-antimatter asymmetry?*
5. *How are the unique properties of QCD manifested in unusual properties of strongly interacting matter?*

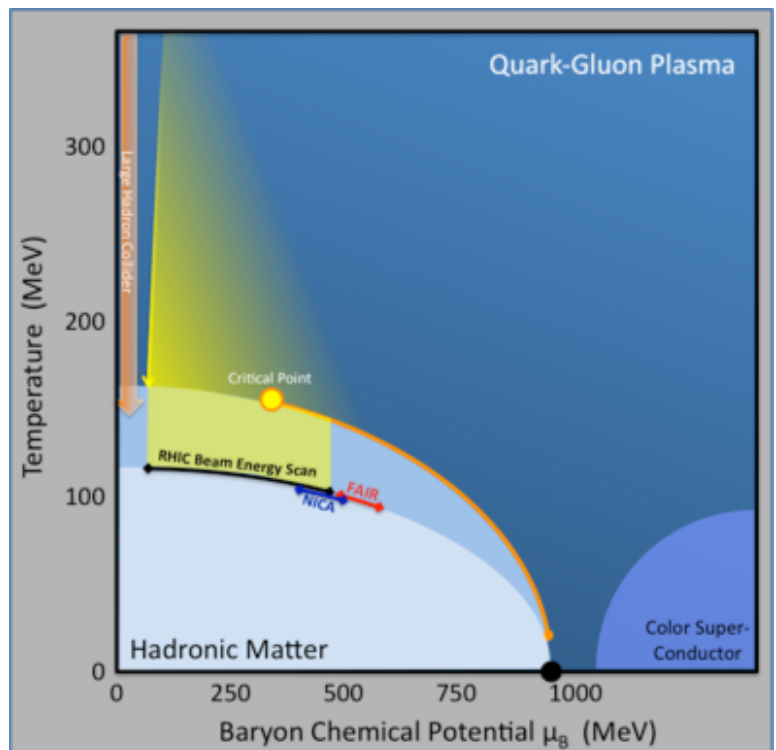

**Figure 1 - Anticipated phase structure of hadronic matter**

Gauge theories describing other fundamental interactions raise similar questions, and presumably the phase structure of each played its role in the very early development of the Universe. The case of QCD is special, because it can be studied experimentally. Based on the results of lattice gauge theory simulations, the transition from hadrons to a quarks-gluon plasma occurs at T ~ 160 MeV, which is low enough to be studied *in the laboratory.* (This is not the case for the electroweak gauge theory, where symmetry breaking occurs at T ~ 100 GeV, out of experimental reach.)

The experimental approach to the study of strongly interacting matter at very high temperatures involves collisions of hadrons at very high energy. The study of the properties of dense, many-body systems– implicit in most of the questions above – requires a large volume of high energy density matter. Thus, the fundamental questions about the physical states of strongly interacting matter are addressed in the laboratory by studying heavy ion collisions at relativistic energies.

In order to gain new understanding of QCD from the interaction of relativistic heavy ions, one needs directly comparable data sets from systems of various sizes, different energies and different experimental probes. These ancillary data sets provide "baselines" against which the largest volume, highest temperature, reactions can be compared. Thus a fundamental and systematic study of QCD requires data on nucleus-nucleus, proton-nucleus (or deuteron-nucleus), as well as proton-proton collisions, all at comparable nucleon-nucleon center of mass energies, and preferably in the same detector systems. Hard processes, which can be accurately calculated in proton-proton collisions using perturbative QCD, can then serve as calibrated probes of the medium created in collisions involving nuclear beams. Addressing many of the questions above also requires data from high energy interactions of hadrons with non-hadronic probes, e.g., deep inelastic scattering (DIS) of leptons from nuclear targets ranging from protons to heavy ions. Addressing the question on the properties of strongly interacting matter also requires the use of spin-polarized beams and/or targets at the highest energies.

Fixed target and collider experiments have contributed to the experimental attack on these questions over time. Taking beam energies of 1 GeV or greater, one has:

- fixed target heavy ion experiments (CERN-SPS, SIS-GSI/FAIR);
- fixed target proton-nucleus studies (Fermilab, J-PARC);
- heavy ion collider experiments (BNL-RHIC, CERN-LHC and in the future NICA);
- fixed target lepton DIS experiments (CERN, JLab-CEBAF);
- DIS collider experiments (formerly at DESY-HERA – in the future possibly at EIC (eRHIC or JLEIC) or LHeC);
- polarized beams (RHIC, future possibilities at eRHIC and JLEIC).

Recent results in this field have primarily come from RHIC at BNL and LHC at CERN. The experiments at RHIC and LHC have produced many new and often unexpected results, which can be summarized as follows:

At nucleon-nucleon center of mass energy $\sqrt{s_{NN}} > 100$ GeV, central collisions of heavy nuclei produce a system that reaches a temperature of approximately 300 MeV (~$4\times10^{12}$ K) and very small baryon chemical potential. This temperature is well in excess of the critical transition temperature predicted by lattice gauge simulations ($T_{crit}$ ~ 160 MeV). A new state of matter is produced under these conditions and observed to have the following properties:

- The matter is an almost "perfect" liquid of quarks and gluons with a shear viscosity-to-entropy density ratio near the quantum limit. This is deduced from the systematics of the collective flow imprinted on the emitted particles, which is well described by nearly inviscid hydrodynamics. Valence quark scaling of the flow indicates that the matter is initially composed of individual quarks, not hadrons. This leads to the conclusion that the produced hot matter is a strongly coupled Quark-Gluon Plasma (sQGP).
- Collective flow effects are also observed for small collision systems at high collision energy, when the number of produced particles is large.
- The matter is opaque to strongly interacting particles (deduced from "jet quenching" measurements) but transparent to real and virtual photons (deduced from direct photon and lepton pair measurements).
- The production of heavy quarkonium states is strongly suppressed when heavy quarks are rare, consistent with partial screening of the color force between heavy quark-antiquark pairs inside the hot matter, but becomes enhanced when heavy quarks are produced abundantly, probably due to final-state recombination.
- Rapidity distributions of produced particles and the suppression of correlated particle production at forward angles in collisions of nucleons with heavy ions are consistent with the existence of a soft gluon component in nuclei called a Color Glass Condensate (CGC).

New questions about the properties of the sQGP have emerged from these discoveries:

1. How close is the shear viscosity-to-entropy density to the quantum bound and how does its value change with temperature?
2. Do heavy quarks (charm, bottom) participate in the collective flow of the sQGP?
3. How small can a droplet of sQGP be and behave collectively as "matter"?
4. Are there quasi-particles in the sQGP that survive at $T > T_{crit}$?
5. What is the color screening length in the sQGP?
6. What are the dominant parton energy-loss mechanisms in sQGP?
7. Is there a critical point in the QCD phase diagram?
8. Is chiral symmetry restored at $T > T_{crit}$?

These and many other questions are under active experimental investigation in a diverse program that includes the study of jets and γ-jet correlations, heavy quarks and quarkonia, low-energy beam scans and proton-heavy ion collisions.

A luminosity upgrade for low beam energies using electron cooling will extend the reach of RHIC into the baryon dense region of the QCD phase diagram where there are indications that the critical point may be located. The CERN-LHC heavy ion program continues with upgrades of all detectors planned in 2019-20. A focus of the experiments at the LHC will be hard probes, such as jets, heavy quarks and quarkonia. These processes, which are produced at the early stages of the collision, are sensitive probes of the collision dynamics at both short and long timescales.

After 2025 complementary studies of the structure of strongly interacting matter are planned at the Facility for Antiproton and Ion Research (GSI-FAIR), where fixed-target ion-ion collisions at lower energies can create nucleus-sized volumes of lower temperature, high baryon density samples of nuclear matter.

Much of our present understanding of the spin structure of strongly interacting matter comes from deep-inelastic scattering measurements of leptons on fixed targets. However, polarized hadron-hadron collisions are directly sensitive to the gluons and over the last several years significant progress has been made in understanding the role of gluons in the proton's spin. Measurements at RHIC have established that gluons make a substantial contribution to the spin of the proton. Spin asymmetry measurements of W boson production in polarized proton-proton collisions at $\sqrt{s}$ = 500 GeV at RHIC provide flavor-separated quark and antiquark helicity distributions.

At high energy, two fundamental aspects of the nucleon partonic structure will remain a focus: One is the nature of the nucleon spin; the other is the nature of the quark and gluon momentum and spatial distributions in the nucleon. There are plans at BNL and Jefferson Lab to open up a new window on deep inelastic lepton-hadron experiments using electron-ion collisions. At BNL this would involve the addition of an electron ring to one of the existing hadron rings. At Jefferson Lab there are plans to add a figure-eight shaped hadron ring to the existing 12 GeV electron accelerator. Both proposals would make available electron-ion and polarized electron-polarized proton collisions and enable studies of the structure of strongly interacting states of matter using precision (lepton) probes in novel kinematic domains. Consideration has also given to adding an electron ring (LHeC) to the CERN-LHC in order to study electron-nucleus collisions at extremely small values of Bjorken-*x* and high luminosity.

**Updated December 27, 2017 by Jens Erler (UNAM)**

With the recent discovery at the LHC of a particle whose properties are fully consistent with those of the Standard Model (SM) Higgs boson, we are about to close a chapter in our quest to understand the fundamentals of the observables universe. The SM, the theory for the strong, electromagnetic (EM) and weak interactions, has to be seen as an overwhelming success, and is in remarkable agreement with the available experimental data (some anomalies aside).

While the details of the SM, such as its particle content and the values of its parameters, are rather ad hoc, its basic structure is dictated by the axioms of quantum mechanics and Lorentz invariance (the independence of physical observables from space rotations and boost transformations) implying an effective field theory (EFT) picture including an ordering principle in terms of mass scales. The fundamental SM Lagrangian arises here as the most general possibility consistent with the assumed SM particles and gauge interactions, and — most importantly — predicts fundamental symmetries such as baryon number (B) and lepton number (L) conservation to leading "renormalizable" order, i.e., up to order four in mass dimension. Various other symmetries are not exact at this level, but their violations are strongly suppressed or are otherwise known to be very small. Thus, high precision tests of fundamental symmetries in particle, nuclear, hadronic and atomic physics serve as indispensable alternatives to the energy frontier, often probing scales not accessible at any existing or planned high-energy collider.

The EFT picture fails dramatically in two respects. The cosmological constant (the unique dimension zero term and possibly the source of the "dark energy" responsible for the accelerated expansion of the universe), and likewise the bilinear (dimension two) Higgs mass term, introduce hierarchies of scales (relative to the Planck scale) which are not understood and moreover are unstable under radiative corrections. It is well possible that at least the problem related to the Higgs may be solved by the discovery of new physics beyond the SM (BSM) with a characteristic scale around a TeV. In this context it is interesting to note that the matter as described within the SM only amounts to 5% of what constitutes the universe. The "dark matter" (about 27%) is possibly a manifestation of TeV scale BSM physics, but may also be associated with very different energy scales. Nuclear physics experiments provide a special quantum context in which selection rules can be used to extract specific components of the new physics with enhanced symmetry violation effects.

For example, one can use the fact that the EM interaction is invariant under space reflections, i.e., under parity transformations (P). With the exception of the tiny $\theta_{QCD}$-term, the gauge theory for the strong interaction (QCD) also respects P, so that experiments measuring parity violating (PV) effects in atomic physics or in polarized electron scattering directly probe the weak interaction and new physics. Note, that one generally expects the latter to be chiral and therefore P violating, since this would shield new fermions from receiving ultra-heavy masses, in much the same way as the SM fermions are massless before electroweak (EW) symmetry breaking (at $\Lambda_{EW}$ = 246 GeV). This strategy is particularly fruitful whenever the SM contribution to a given PV observable is small, which is the case for the left-right polarization asymmetry in both Møller (E158 at SLAC and MOLLER at Jefferson Lab) and ep scattering (Qweak at Jefferson Lab and P2 in Mainz). These experiments will determine the weak charges ($Q_W$) of the electron and the proton, respectively, and will not only probe multi-TeV energies, but also test the scale ($\mu$) dependence ("running") of the central EW gauge parameter, the weak mixing angle, $\sin^2\theta_W$, as illustrated in the figure.

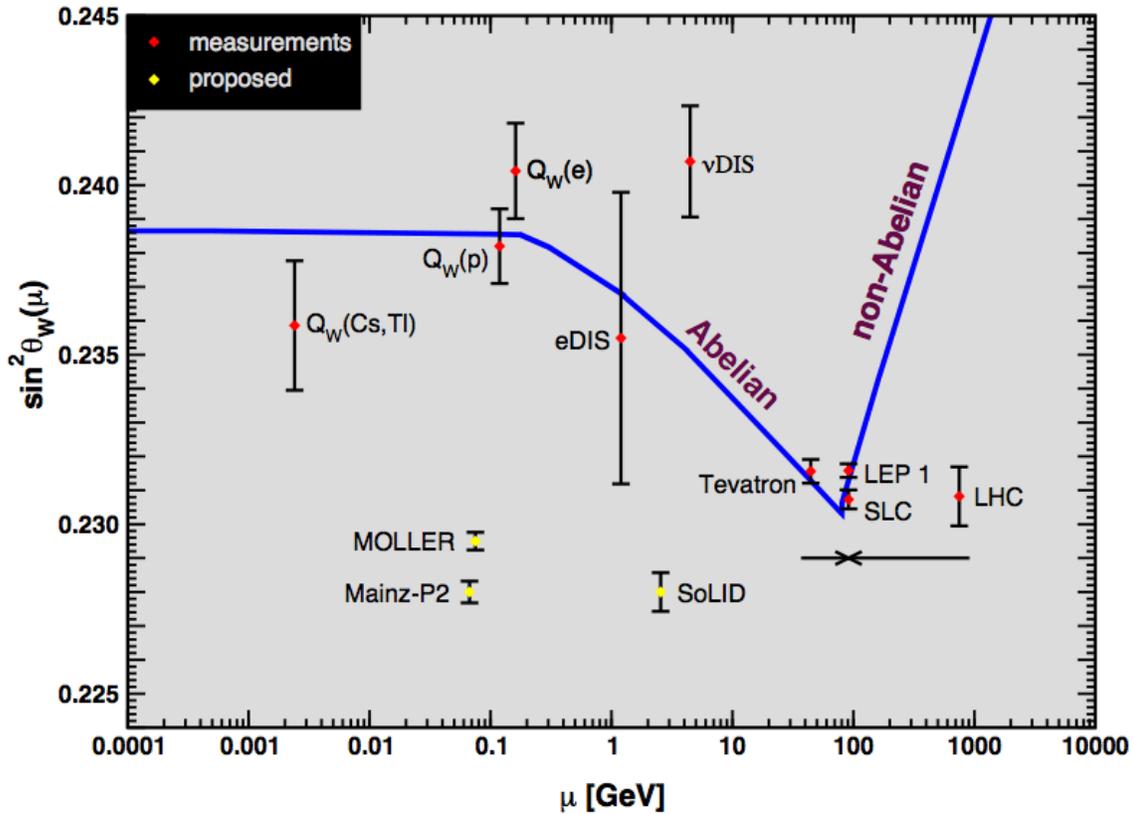

*Calculated running of the weak mixing angle in the SM, as defined in the modified minimal subtraction scheme. The theoretical uncertainty is below the thickness of the blue curve. Red points with error bars show existing data and yellow points (with arbitrarily chosen ordinates) refer to future experiments. $Q_W(Cs,Tl)$ derives from atomic parity violation, $Q_W(p)$ from Qweak, and $Q_W(e)$ from SLAC-E158, while vDIS and eDIS denote deep inelastic neutrino (NuTeV at Fermilab) and electron scattering, respectively. However, the interpretation of the former is obstructed by nuclear effects and by quark charge symmetry violation effects, which need to be understood independently before electroweak physics can be extracted unambiguously. The eDIS point is inferred from scattering off deuterons which is easier to treat theoretically and which is dominantly from the recently completed PVDIS experiment at the 6 GeV CEBAF at Jefferson Lab. SoLID is with a future detector to be exposed to the upgraded (12 GeV) beam for a more precise follow-up experiment. LEP 1, SLC, Tevatron, and LHC, all refer to weak mixing angle extractions from Z pole asymmetries (for clarity the Tevatron and LHC points are shown shifted horizontally). Another possibility (not shown in the figure) is to map out the sub-Z pole region to about % precision at an electron-ion collider (EIC). Likewise, ep modes of the LHC (LHeC) and of other possible future circular colliders like the FCC (CERN) or the CEPC/SppC (China) would be able to populate the above-Z pole branch.*

This class of experiments can also constrain the properties of a dark photon which has been hypothesized as the mediator between a dark sector (possibly responsible for the dark matter) and our visible world. Interestingly, such an object could also affect the anomalous magnetic moment of the muon, $g_\mu-2$, which is known to deviate at the level of almost four standard deviations from the SM. Thus, it is of utmost importance to improve the experimental precision in $g_\mu-2$ which is the goal of new experiments at FNAL and J-PARC, and to reduce the theoretical (hadronic) uncertainties in its SM prediction.

Searches for dark photons and other very weakly coupled new light particles are part of a larger effort to discover the nature of the dark matter. For example, direct detection experiments are in full swing trying to detect the scattering of hypothetical weakly interacting massive particles (WIMPs) with nuclei. And the search for new sub-millimeter range forces is well motivated theoretically and may provide a link to both dark matter and dark energy.

One may also look for new Lorentz structures which are absent in the SM at leading order. The charged current weak interaction takes the form of a specific combination of vector and axial-vector (V−A) terms (as tested in many β decay experiments) and hence exhibits maximal PV. However, the present set of experimental data cannot exclude the presence of scalar, pseudoscalar, tensor or V+A terms at the few percent level. It is then one of the goals of the program of testing fundamental symmetries to tighten these constraints on the Lorentz structure of the weak interaction through semi-leptonic decays (nuclear decay distributions) and purely leptonic decays of muons (such as by the TWIST Collaboration) and taus. Related approaches attempt to test lepton universality to very high precision in decays such as $\pi \to l\, \nu_l\, (\gamma)$ with l = e or µ (at PSI and TRIUMF) or to study the predicted unitarity of the CKM mixing matrix connecting the mass and interaction eigenstates of quarks. In particular, there is a worldwide effort to determine the CKM element $V_{ud}$ from neutron decay to high precision to match or surpass the $V_{ud}$ precision from super-allowed Fermi nuclear β decays. Nab at the SNS, PERC in Munich, as well as PERKEO III and UCNA will improve the constraint on the axial-vector Gamow-Teller transitions, and the existing experimental discrepancies in the neutron lifetime derived from cold neutron beams on the one hand, and ultra-cold neutron storage experiments on the other, need to be resolved.

Much higher energy scales can be probed by electric dipole moments (EDMs). They violate time reversal invariance (T) which in any quantum field theory (QFT) is equivalent to CP invariance — the product of charge conjugation (C) and P. The observed CP violation (CPV) in K and B-meson systems is fully accounted for by the complex phase ($\delta_{CKM}$) appearing in the CKM matrix. However, $\delta_{CKM}$ cannot induce effects in EDMs that would be large enough to be detected in any current or planned experiment. As a consequence, if a permanent EDM was observed it would be tantamount to the discovery of a BSM effect with very far-reaching consequences. To understand the deeper origin of the effect it would then be necessary to experimentally probe EDMs in as many systems as possible, including leptons, nucleons, nuclei, diamagnetic and paramagnetic atoms, and molecules. Since the $\theta_{QCD}$-term could induce a nucleon EDM, it would be particularly interesting to isolate a leptonic EDM. The current limit for the electron EDM, $d_e$, by the ACME Collaboration can be expressed as $|d_e| < e\, \Lambda_{EW}\, (227\text{ PeV})^{-2}$ (with e the fundamental electric charge) which sets the sensitivity scale. More than half a dozen experiments worldwide are aiming to strengthen the limit on the very complementary neutron EDM to a similar level and many other systems are being explored. The search for new sources of CPV should be a priority in that the observed baryon-antibaryon asymmetry in the universe (BAU) can only be explained by CPV beyond $\delta_{CKM}$, and also since most BSM scenarios introduce many new complex CPV phases.

Such enormous scales can also be reached in charged lepton flavor violation experiments, including $\mu^+ \to e^+ \gamma$, $\mu^+ \to e^+\, e^-\, e^+$, $K_L \to \mu^\pm\, e^\mp$, muonium-antimuonium oscillations and µ to e conversion. The complementarity of these processes in the context of different BSM scenarios is well appreciated. Furthermore, the strong suppression of flavor-changing neutral currents in the SM allows for comparable sensitivities in decays such as K → π ν ν̄. Future measurements of the charged (CERN) and neutral (J-PARC) modes will also provide unique constraints on the smaller elements of the CKM matrix.

The BAU also calls B conservation into question. Proton decay experiments already ruled out simple scenarios of Grand Unified Theories of the strong and EW interactions, probing scales in the vicinity of the fundamental Planck scale. The Planck scale itself may come into play should the QFT framework break down, signaled, e.g., by a violation of CPT invariance. The most sensitive tests look for differences in the masses or lifetimes of particles and antiparticles or compare the atomic spectra of hydrogen with antihydrogen as was recently achieved by the ALPHA Collaboration at CERN.

Neutrino mass and mixing are now believed to be at least the major contributor to the phenomenon causing neutrinos to oscillate between flavor eigenstates as observed for solar neutrinos, atmospheric neutrinos, reactor antineutrinos, and accelerator neutrinos. The corresponding Pontecorvo-Maki-Nakagawa-Sakata (PMNS) matrix is completely analogous to the CKM matrix, except that it allows for two additional complex CPV "Majorana" phases provided the neutrinos are Majorana particles (their own antiparticles). One can incorporate Majorana ν masses within the SM (without introducing new particles) but only if one includes dimension five terms into the EFT of the SM. In this picture, the neutrino mass scale of order 100 meV or less would be generated by a very large "see-saw scale" roughly of order $10^{14}$ GeV. Majorana mass terms violate L conservation and would generate ν-less double β decay. This would be a fundamental and new kind of process, and would also determine the absolute ν mass scale (provided that the transition nuclear matrix elements can be evaluated with sufficient accuracy), while ν oscillations are only sensitive to mass-square differences. Currently, the most stringent limit on the lifetime of $5.3 \times 10^{25}$ years against ν-less double β decay has been achieved by the GERDA Collaboration. The more traditional way to determine the absolute ν mass scale is by measurements of β decay spectra near their kinematically allowed endpoints, such as the new tritium decay experiment KATRIN with a projected sensitivity to $ν_e$ masses down to 200 meV. Moreover, the Project 8 Collaboration demonstrated the viability of a technique for β spectroscopy based on cyclotron radiation which would allow for even greater sensitivity.

On the other hand, should L be a fundamental symmetry of nature one would need to introduce right-handed neutrinos to allow for Dirac masses just as for quarks (but with tiny Yukawa couplings). It is stressed that with the recent extraction of the smallest of the three mixing angles ($θ_{13}$) the reactor experiments Double Chooz, Daya Bay and RENO have completed the determination of the mixing part of the PMNS matrix. Moreover, $θ_{13}$ is large enough that one can hope to discover a non-trivial value of the CPV Dirac phase (the PMNS analog to $δ_{CKM}$) by studying the difference between neutrino and the corresponding antineutrino oscillations in experiments based on long baselines. Indeed, first results from the T2K experiment provide a statistically weak indication (at the level of about two standard deviations) for a CPV Dirac phase in the PMNS matrix, joined by an even weaker preference for a ν spectrum with a normal hierarchy, i.e., a spectrum in which the lightest mass eigenstate has the greatest overlap with the electron neutrino and which thus mirrors the quark hierarchy. Further insight may be gained from experiments studying neutrinos traversing the Earth to isolate matter effects. Some experimental anomalies in the neutrino oscillation sector may be interpreted in terms of "sterile" ν states (extra gauge neutral fermions) calling for further experimental efforts to clarify our understanding.

**Updated January 1, 2018, by Marco Durante (TIFPA-INFN)**

Over the past five years, applications of nuclear physics continued to grow and to produce benefits for the society. The latest Long Range Plans of the Nuclear Physics European Collaboration Committee (NuPECC; 2017) and of the US Nuclear Science Advisory Committee (2015) identify several specific applications that characterize the broader impact of nuclear sciences in the society. Recent developments in the main applications are summarized below.

*Medical applications*

Applications of nuclear physics in medicine are perhaps those with the largest expansion potential in nuclear physics. According to the 2014 NuPECC report "*Nuclear Physics in Medicine*", medical applications can be further sub-divided into three sectors.

Imaging

Radiography was the first medical application of X-rays and also today represents the most obvious benefit of nuclear physics in medicine. Recent improvements include spectral CT, such as K-edge imaging, based on hybrid pixel detectors; 3- photon cameras (gamma-PET); and ultra-high field MRI. Even if not yet used in clinics, 7T MRI will provide anatomical detail at the submillimiter scale, enhanced contrast mechanisms, outstanding spectroscopy performance, ultra-high resolution functional imaging (fMRI), multinuclear whole-body MRI and functional information. Industrial developments of CT scans are also remarkable, and the most recent devices can scan at 737 mm/s, thus making a chest scan in less than a second and a full body scan in five seconds only.

Particle therapy

Radiotherapy with accelerated charged particles, a typical medical application of nuclear physics, is rapidly growing. At the end of 2017, there are 63 centers worldwide for cancer treatment with protons (52 centers) or carbon ions (11 centers). It is estimated that over 200 centers will be active by 2021. However, despite increasing evidence of clinical efficacy and low toxicity, the method remains controversial. In fact, particle therapy remains much more expensive than X-ray therapy. The success of this therapy in the future will strongly depends on progress in nuclear physics leading to a reduction of the costs – especially the construction of smaller, compact accelerators and beam delivery systems (gantry). Moreover, it will be necessary to improve the benefit of the treatment, i.e. to show that it can provide a decisive advantage also for patients not traditionally elected for particle therapy (i.e. all pediatric patients and adult patients affected by chordomas, chondrosarcomas, bone and soft tissue sarcomas, ocular tumors). Several comparative clinical trials are ongoing to prove the superiority of charged particles over X-rays for high incidence and mortality tumors, such as pancreas cancer. Results of these trials are necessary to demonstrate that the physical and radiobiological rationale of particle therapy translates into clinical benefit. In recent years, a major breakthrough in cancer therapy has been the success of immunotherapy for treatment of highly lethal cancers, especially melanoma. Excellent results have also been obtained with targeted cancer therapy. The combination of radiotherapy with these new drugs is currently the main topic of interest in oncology, and the role of charged particles in this frame has just started to be tested. Another rapidly expanding field is that of noncancer diseases. Stereotactic body radiotherapy with X-rays has been successfully applied for the first time to the treatment of ventricular arrhythmia. Pre-clinical studies using carbon ions in a swine model suggest that particle therapy may cure heart arrhythmia with reduced toxicity (Figure 1). This application may largely increase the number of patients who can benefit from particles.

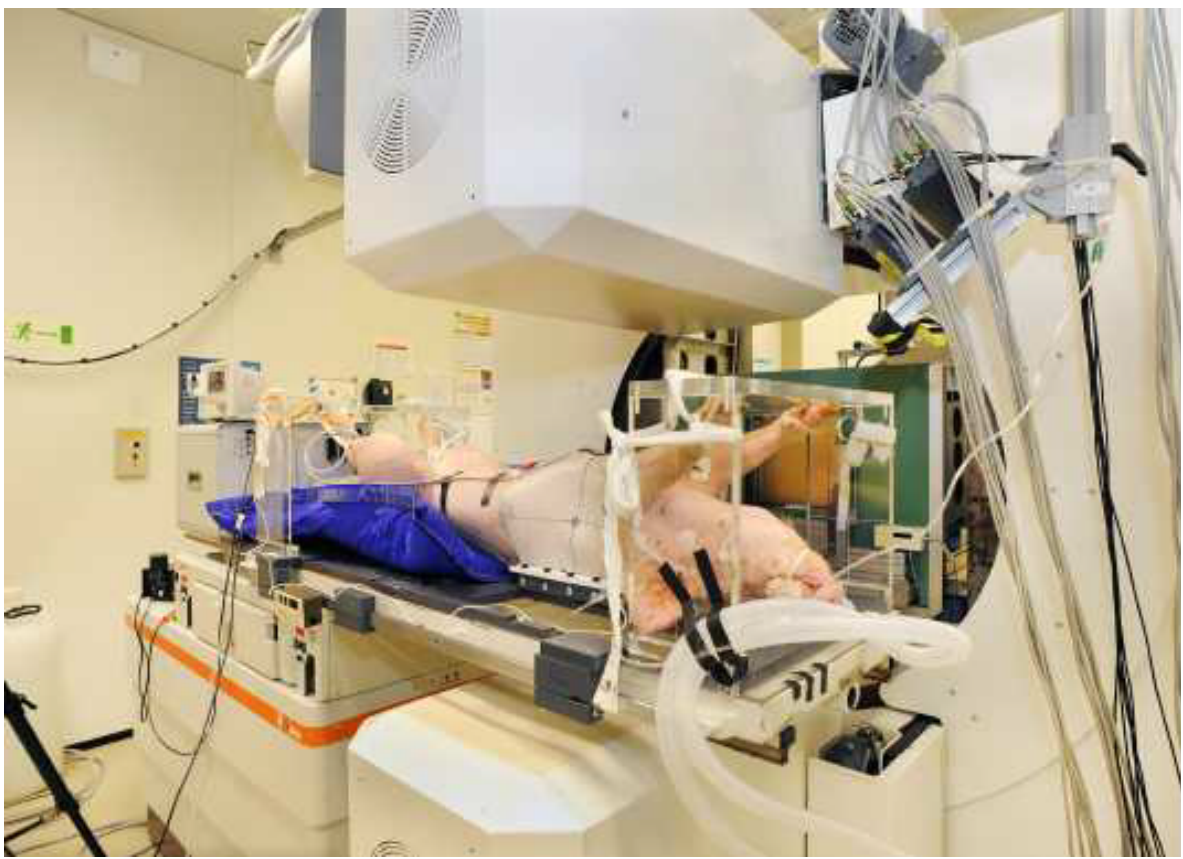

**Figure 1 - Particle therapy for heart disease.** An anesthetized pig during an irradiation with high-energy 12C-ions at the GSI Helmholtz center in Darmstadt (Germany). The beam, coming from the orange window, is directed to the heart of the animal in a feasibility study for cardiac arrhythmia ablation using external, non-invasive, heavy ion beams. Online PET imaging is provided by the detectors on top and bottom (Lehmann et al., Sci. Rep. 6:38895, 2016)

Radioisotopes

The use of radionuclides for diagnostic purposes (SPECT and PET) is a mature field in nuclear medicine. The main problem is the shortage of the common isotope $^{99}$Mo/$^{99m}$Tc (emitting a 140 keV photon with a half-life of approximately 6 hrs), the "workhorse" for imaging by gamma cameras and SPECT. New facilities based on fission reactors, cyclotrons, or high intensity linear accelerators are necessary to ensure a regular supply of these important radioisotopes. Therapeutic applications of the radionuclides have been growing in the past years. The use of $^{131}$I for thyroid cancer has been established for many years, but the recent introduction of new vectors (antibodies, peptides, and folates) makes possible to deliver radionuclides to many common, resistant cancers, and to target micro- metastasis or minimal residual disease following surgery or teletherapy. For targeting single cells, isotopes emitting $\alpha$-particles or Auger electrons are interesting, because they have short range and high biological effectiveness. A few clinical trials are already ongoing, but the main limitation remains the availability of the exotic isotopes (e.g. $^{177m}$Sn for Auger or $^{225}$Ac for $\alpha$-particles). New interesting applications are those in the new field of theranostics, where radioisotopes are used simultaneously for imaging and treatment. This requires the combination of two, chemically identical (e.g. $^{64}$Cu/$^{67}$Cu) or similar ($^{99m}$Tc/$^{188}$Re) isotopes, bound to the same vector; or a single isotope (e.g. $^{177m}$Sn) able to destroy the target cells and to be visualized externally.

*Environmental applications*

Nuclear physics has greatly contributed to our understanding of climate changes and the impact of anthropogenic activities. Proton–induced X-ray emission (PIXE), Ion Beam Analysis (IBA), and Accelerator Mass Spectrometry (AMS), can accurately measure the composition of the atmospheric aerosol and give to policymakers the knowledge and the tools for a significant reduction in anthropogenic emissions. The new US administration reverted previous environmental policy pulling out of the Paris climate agreement. The development of efficient technologies for elemental and radionuclide analysis to monitor environment changes is therefore becoming more and more relevant. Studies of environmental radioactivity also rapidly evolved in the past few years. Beyond the radioactive waste of nuclear power plants, the focus is now moving toward naturally occurring radioactive materials generated in oil, gas, and mineral production industries, including fracking. Notwithstanding the European efforts in funding large research programs (the EU MELODI platform), large uncertainties remain on the low-dose radiation risk. The discovery of the bystander effect and evidence of late risk of noncancer diseases (especially cardiovascular mortality) may indicate a supralinear risk at low doses. On the other hand, lack of increased morbidities in high-background radiation areas (such as Kerala in India or Ramsar in Iran) suggests a sublinear risk at low doses. Interesting new radiobiological experiments planned in underground particle physics laboratory (INFN Gran Sasso in Italy and SNOLAB in Canada), designed to have zero-background, will help understanding the effect of background environmental radioactivity on living organisms.

*Space radiation*

The plans of Space Agencies have recently moved from the International Space Station in Low Earth Orbit (LEO) to manned exploration beyond low-Earth orbit (BLEO), especially to the moon, asteroids, and finally Mars. For exploration in deep space, it is generally acknowledged that exposure to cosmic rays represents the main health hazards for the crew. The recent measurements of the NASA RAD detector on the Mars Science Laboratory during its trip to Mars (11.6.2011 –8.6.2012) indicate that the dose rate in BLEO is about 1.8 mSv/day. This means that in a mission to Mars, crewmembers can absorb a dose close to 1 Sv. This is a very high dose and, even if precise dose limits for BLEO have not been issued by the Space Agencies, it would not be allowed in any terrestrial activity. Accelerator- based research programs have been established both by NASA (at the Brookhaven National Laboratory in NY) and ESA (at GSI in Darmstadt, Germany, and in the future at the new FAIR facility) to reduce uncertainty on the risk and to develop countermeasures. Passive shielding remains indeed the only practical countermeasure available, but the development of new, highly hydrogenated multi-functional materials is necessary, and tests at high-energy heavy ion accelerators are used to assess the shielding effectiveness (Figure 2). Future mitigation strategies may include active shielding (with superconductive magnets) and radioprotective drugs or dietary supplements.

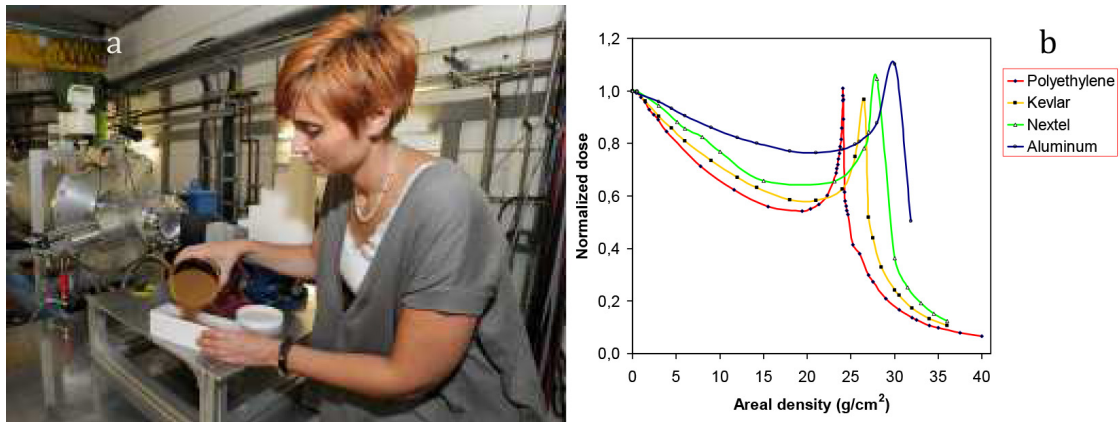

Figure 2 - Space radiation shielding tests. Accelerator tests of the shielding effectiveness of different space relevant materials. (a) Mars regolith is prepared at the GSI accelerator in Germany. (b) Measured Bragg curves of 1

*Societal applications*

New technical developments have strengthened the position of nuclear techniques compared to other methods for the benefit of the society in many different fields. A few examples are listed below.

Cultural heritage

IBA has been used for decades to analyze art objects. The main strength is its analytical performance, which reaches the trace elemental level, without sampling. Although IBA is considered non-invasive, it has been recently demonstrated that it can induce visible, irreversible changes, depending on the material and on the technique. One of the obvious mitigation strategies is to decrease the beam current and the time of acquisition using more efficient detector systems. The AGLAE Laboratory in the Louvre museum has managed to gain a factor of ten for trace elements analysis with their new detector configuration. Among the new techniques under study, laser-accelerated protons may in the future replace electrostatic accelerators for PIXE, and high-energy monochromatic ɣ-rays are very promising for imaging of large objects. Radiography and tomography using very intense ɣ-ray beams of small bandwidth and high energy will allow high-resolution 2D/3D imaging and in-depth elemental analyses of large objects of various nature and composition.

Archeometry

Radiocarbon dating is the classical method for dating of archaeological specimens. Carbon-14 measurements by AMS allow dating objects from tens of thousands of years ago to the middle of last century with an accuracy of few tenths of years. However, detonation of the atmospheric nuclear bombs during the Cold War years, prior to the Test Ban Treaty in 1963, increased the amount of radio-carbon in the atmosphere significantly, providing a time marker. The so-called "bomb- peak dating" addresses the interesting period of the last sixty years with an accuracy of about two years. Thanks to the improved sensitivity of AMS, it has then be possible to measure the age of the cells within a human body, including human cardiomyocytes, adipocytes, oligodendrocytes and neurons. The method also finds many applications in forensic science, because it can be sued to date the age of death in corpses, and cultural heritage, helping identifying forgery of modern art works (box 3).

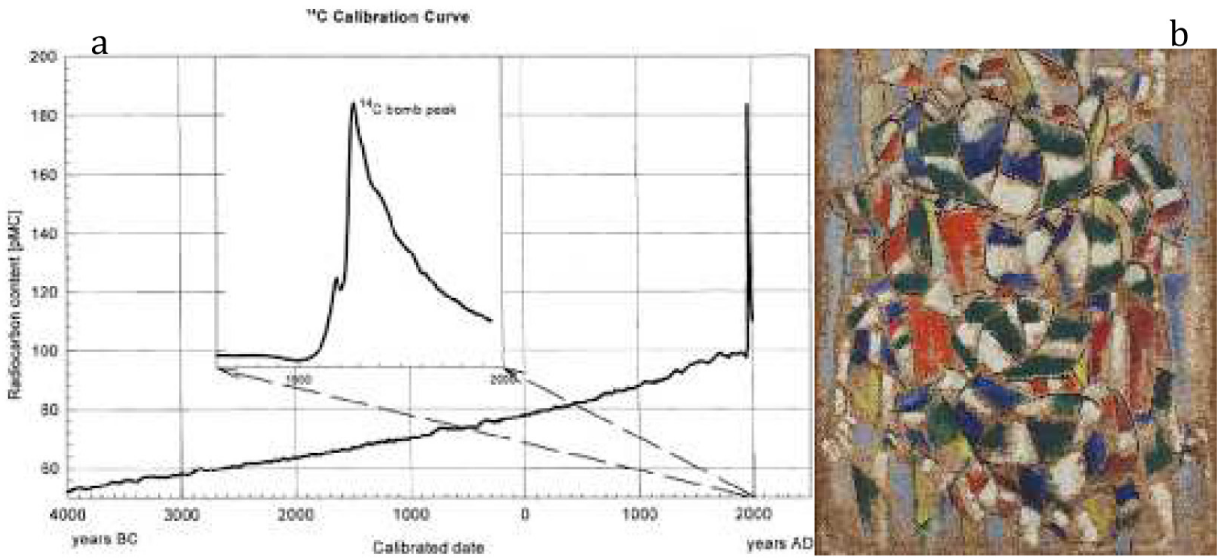

**Figure 3 -  Bomb-peak dating. Nuclear atmospheric tests in the 60s' have modified the classical 14C calibration curve, opening a window of opportunity for high- resolution dating of objects in the past 60 years. (a) The excess is now almost back to the backgr**

Homeland security

The possibility that terrorist organizations will use chemical, biological, radiological or nuclear weapons for future attacks is considered credible and highly probable. Several attempts have been prevented by security services. Nuclear physics can play an important role in prevention of nuclear and radiological terrorism. Screening for illegal transportation of radioactive material is generally performed by radiation portal monitors. For plutonium trafficking, $^3$He sensors can detect a characteristic neutron emission. However, shielding and background radiation hamper the effectiveness of these detectors. In recent years a large number of new high light-yield scintillator materials have been discovered. In particular, Lanthanum halides provided the starting point for the design and development of several new high performance detector arrays. Muon radiography is also considered a promising technique for searching high-Z materials in trucks. Active methods using small accelerators would be more accurate and sensitive. For instance, a 3 MeV deuteron beam can be used to induce the $^{11}B(d, n\gamma)^{12}C$ reaction, and the different $\gamma$-ray peaks provide independent information on the density and the atomic number of the material (Box 4).

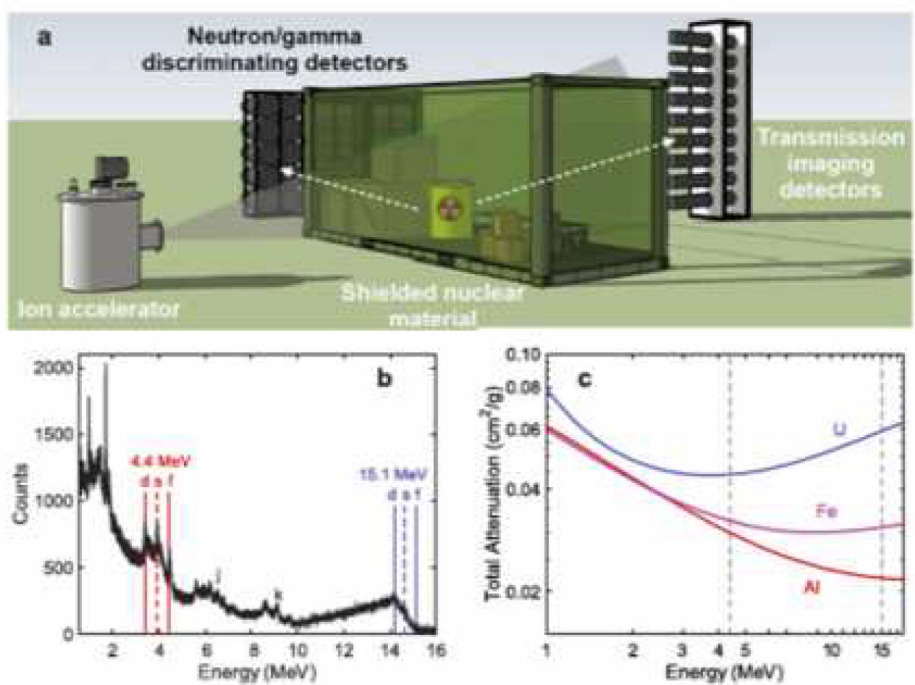

Figure 4 - Detecting illicit weapon trafficking. (a) Low-energy nuclear reaction imaging relies upon the source of monochromatic photons via a nuclear reaction between an ion accelerated to MeV-scale energy and a target. Gamma rays at discrete energies are produced from nuclear excited states of the product nucleus, with some reactions also producing neutrons. The collimated, penetrating radiation from the nuclear reaction source is used to perform transmission radiography of a shielded object, while neutron/gamma discriminating detectors detect the signature of nuclear fission. (b) Photon spectrum from the 11B(d,nγ)12C reaction measured with a LaBr scintillation detector. (c) Energy-dependent attenuation for several elements. The 4.438 MeV and 15.1 MeV gamma energies from the 11B(d,nγ)12C reaction are shown as dashed lines (Rose *et al.*, *Sci. Rep.* 6:24388, 2016).

*Conclusions*

Nuclear physics provided society with a large number of beneficial applications in many completely different fields including medicine, heritage, and security. In the past few years, large efforts have been dedicated to technological improvements, especially in radiation detectors, able to enhance these applications. Future improvements in accelerator design, leading to smaller, compact and cheap particle accelerators, will further boost many applications. In addition, nuclear physicists are recognizing the value of inter-disciplinary research, and very often work closely together with chemists, engineers, biologists, or physicians in all the different steps of the application. Applied nuclear physics is growing as a large and powerful branch of nuclear sciences, bringing benefit to the society and breakthrough research results in other scientific disciplines.

# Future Nuclear Physics Facilities around the World

**Updated January 1, 2018, by Hideto En'yo (RIKEN, Nishina Center for Accelerator based Science)**

This summary is updated from the original manuscript which was published in IUPAP Report 41, dated August 12, 2013. During the last four years, striking updates happened in the world. The Russian NICA project was approved as a medium energy heavy-ion and polarized-proton collider to be built in Dubna. The Chinese Heavy Ion Accelerator Facility (HIAF) project, was approved to be built at Huizhou, Guangdong Province (100km north-east of Hong Kong) as a nascent of the Heavy Ion Research Facility in Lanzhou (HIRFL). Next to HIAF, the China Accelerator Driven Sub-critical System (CIADS) will also be built. In Europe, The scope of the German FAIR project was redefined and a clear project schedule has been drawn up.

In the coming 10 years, together with these "new" comers, SPIRAL2 will take off in France, FRIB in the USA, RAON in Korea, and FAIR in Germany and many other new facilities currently under construction will come online. This is a surprisingly superb situation for the world nuclear physics community and such a situation will bring a quantum jump in the research field of nuclear science.

On the other hand, however, some of these facilities have reached a stage that involves a burden of construction (in terms of budget and human-resources) which is beyond a capacity of a single nation. The price tag for a nuclear physics facility may not be as large as projected for the International Linear Collider (ILC), but the sum of the construction costs for all the nuclear physics facilities (currently running or under construction) in the world is reaching the level of construction costs of the ILC.

Clearly this fact needs to be recognized and efforts must be made to enhance worldwide cooperation while keeping a good balance between domestic, regional (Europe-Africa, Asia-Oceania, and North-South America), and truly international projects. IUPAP WG.9 has recognized its mandate in dealing with such international issues and plans to take a more explicit role on international cooperation for the large scale nuclear science projects as also requested by the nuclear science funding agency/government representatives at the Nuclear Science Symposium held at the RIKEN Tokyo Office, August 30-31, 2017. This summary is to provide an overview of the current and future nuclear physics "user" facilities worldwide.

**Table 1**

| Subfield | Scientific Head Lines |
|---|---|
| Quark many body -Hot / Dense QCD- | Quark Gluon Plasma (QGP) -properties of the early universe- |
| Quark many body -Cold QCD- | QCD Chiral Symmetry -nucleon/meson structure and properties- |
| Nucleon many body systems -Nuclear Structure- | Ultimate Nuclear Picture (including hyper nuclei) Element Genesis and Astrophysics Super Heavy Element and the Island of Stability |
| Fundamental Physics | Physics of the Lagrangian Double beta decays, Neutron EDM, rare decays, *etc.* |
| Computation | Ab-initio calculations, Lattice QCD |
| Application | Nuclear Transformation, Catalyzed Fusion, *etc.* |

Table 1 summarizes the subfields of nuclear physics and the main topics to be covered. All the subfields contain prominent scientific topics. This manuscript focused on the facilities which carry out experiments in the first three subfields, namely, "Hot QCD", "Cold QCD", and Nuclear Structure Physics. The latter category requires radioactive (or rare) isotope (RI) beams either from an ISOL (Isotope Separation On-Line) facility or an in-flight projectile fragmentation facility with heavy-ion beams.

Table 2 List of large-scale nuclear-physics facilities. The bold characters are for the running facilities, italics for those under construction, and the gray cells for proposed facilities. The electron-positron colliders are also listed in parentheses.

| Subfield | | America | Europe | Asia |
|---|---|---|---|---|
| Quark many body -Hot QCD- | | **RHIC A+A** | **LHC (ALICE)** *FAIR(SIS100)* *NICA A+A* | |
| Quark many body systems Cold QCD- | hadron beam | | **CERN SPS** *FAIR(SIS100)* | **J-PARC** **HIRFL** |
| | lepton beam | **JLAB-12GeV** | **CERN SPS μ** **MAMI** | **Spring-8** **ELPH** |
| | collider | **RHIC p+p** | *NICA p+p* | **(Bess-III)** **(Belle-II)** |
| | | *eIC e+A/e+p* | | *-eIC@HIAF* |
| Nucleon many body systems | Projectile Fragmentation RI beams | *FRIB* | *FAIR* | **RIBF** **HIRFL** *RISP* *HIAF* |
| | ISOL RI beams | *ARIEL/ISAC2* | **HIE-ISOLDE** *SPIRAL2* *SPES* | *BRIF2* *JUNA* |
| | Super Heavy | **High Flux Reactor** | **GSI UNILAC,** *Dubna SHE factory* | *Superconducting RIKEN RILAC* |
| Super ISOL | | *FRIB upgrade* | *EURISOL* | *Beijing ISOL* |

Table 2 gives the list of the large-scale user facilities that are in operation, are under construction at the present, or are being proposed, in each geographical region. One notices that on the American continent, the projects are well distributed probably due to the control exerted by the US government. The mechanism working on the American continent is a two-body interaction, with Canada cleverly covering the subfields not covered by the US. Europeans are trying a kind of role-sharing by commonly proposing the flagship facilities FAIR and EURISOL together with intermediate steps at distributed facilities in various countries. In Asia, there is no such co-operation. Every nation is advancing its own research interests under the (weekly coupled) three nation entity made up by China, Korea and Japan. The recent establishment of ANPhA (Asia Nuclear Physics Association) may be a key to improve the present situation regarding mutual collaborations in Asia. IUPAP WG.9 may then be responsible for furthering applicable worldwide co-operations.

Hot QCD: This subfield actually started in the late seventies at LBL in the US. In the mid-eighties relativistic heavy-ion programs emerged at the SPS of CERN and the AGS of BNL. Until the very end of the SPS heavy-ion physics program, it was not clear whether the conditions of the Quark Gluon Plasma (QGP) had been reached.

RHIC started operations in 2000 and finally the discovery of a new state of matter (QGP) could be announced. The ALICE experiment at the CERN-LHC started in 2010 and re-discovered this new state of matter. After 30 years of research by increasing the collision energy through five accelerators, convincing demonstrations have finally been obtained regarding the existence of this new state of matter. RHIC and LHC have another 10-15 years of productive research in studying the properties of QGP. It may then very well be that the relativistic heavy-ion colliders are replaced by electron-ion colliders (eIC's). A white paper for such a facility is being presented to the US government and an assessment of the physics with such a facility is currently being made by the US National Academy of Sciences. At CERN there are discussions about a high energy electron-ion collider (LHeC Project).

The history of the "Hot QCD" research has been quite successful not only by reaching in the phase diagram the conditions leading to QGP but also by the healthy growth of international collaborations; researchers got together at accelerators at which one nation alone could not sustain the heavy-ion research program. However, it should be remembered, that none of the facilities for the QGP studies was built from "a green field" by the nuclear physics community (i.e., they were realized by converting an existing accelerator or using an existing tunnel to construct a heavy-ion collider).

What comprises "Dense QCD" is studied at lower-energy heavy-ion collisions, of which FAIR (Facility for Antiproton and Ion Research) at GSI and NICA (Nuclotron based Ion Collider fAcility) at JINR/Dubna are under construction. The FAIR phase-1 scheme is established in 2015 to complete SIS-100 and the experimental facilities by 2025. The first beam circulation of NICA is expected in 2020. Although the energy regions which FAIR and NICA are targeting are the region once covered by the SPS and low-energy scans at RHIC, no experiments were done with high luminosity. FAIR and NICA will challenge the physics of dense QCD at a much higher interaction rate with advanced experimental technology. NICA will also be a polarized proton collider, so it is indeed a mini RHIC.

When FAIR will come on line, this will be the leading European nuclear physics facility covering "Hot QCD", "Cold QCD", and "Nuclear Structure" with RI beams via projectile fragmentation. It is pity that the original double ring structure of SIS100/SIS300 was postponed. Nevertheless the physics program is well thought out, with an emphasis on those experiments where other facilities cannot compete; consequently no neutrino physics experiments, no hyper-nucleus physics experiments.

Cold QCD:  Until FAIR and NICA become operational, J-PARC's 30 GeV Proton Synchrotron in Japan will continue to be the world's leading facility for "Cold QCD" physics using hadron beams. The beam power is now a stably 40kW for the slow extraction and 400kW for the fast extraction, both reaching one half of the original design goal. Recent discovery of a bound K⁻pp system is a highlight.

Amongst a handful of lepton-photon beam facilities in the world, Jefferson Laboratory (JLab) in the US is playing the leading role. The CEBAF energy was upgraded from 6 GeV to 12 GeV, and a new experiment started in 2017 at the Hall D using a Bremsstrahlung-photon beam with a new $4\pi$ detector. The CLAS spectrometer in Hall B was also largely upgraded and engineering runs started also in 2017.

The world's first electron Ion Collider (eIC) is proposed to be built either at BNL by adding an electron ring in the RHIC tunnel or by adding two ion-cooler rings to CEBAF at JLab. The electron-proton colliding luminosity will be 100 times higher than at the former HERA collider. A new regime of QCD, the saturated gluonic field in the nucleon and nucleus will be explored. Similar option is also under consideration at LHC at CERN and HIAF (China).

In addition, one should not forget the hadron physics capability of the electron- positron colliders where many new hadronic resonances have been discovered. BES-III and BELLE-II will continue to be frontrunners in exotic hadron searches.

Nuclear physics with ISOL beams: EURISOL is the future facility with a multi-billion-dollar price tag, carrying the aspirations of the European nuclear physics community. Twenty institutions in Europe signed on to the design report with in addition contributions from twenty other institutions from America and Asia. The Nuclear Physics European Collaboration Committee (NuPECC) has agreed that EURISOL, together with FAIR at GSI are priority goals. Other ongoing projects such as HIE-ISOLDE at CERN, SPIRAL2 at GANIL, and SPES at INFN-Legnaro are now regarded as "intermediate" facilities. This is a very clever and honest approach that satisfies national pride as well as determines the future direction of the projects within the EU.

Although declared as "intermediate", these facilities are anything but minor and are already powerful tools for research. HIE ISOLDE at CERN is a natural extension of the present facility by the addition of a 10MeV/A post-accelerator. SPES at INFN-Legnaro is a relatively modest 8kW-ISOL project. SPIRAL2 in GANIL involves building a 200kW ISOL driven by a deuteron super-conducting linac, which doubles the scale of the present GANIL facility. A low-energy RI beam laboratory and a neutron laboratory are also scheduled to be built.

Table 3 summarizes the ISOL facilities under construction or being planned. While the existing facilities provide $10^{11}$-$10^{12}$ fissions/s with a typical $^{132}$Sn intensity of $10^6$, the European "intermediate" facilities provide $10^{12}$-$10^{14}$ fissions/s with $^{132}$Sn of $10^8$-$10^9$. In Canada, ARIEL at TRIUMF is under construction which aims at $10^{14}$ fission/s with the use of a super-conducting electron linac developed in conjunction with the ILC R&D program. These facilities under construction will provide typically 100 times more RI compared to current levels. For the far future, EURISOL aims at $10^{16}$ fissions/s driven by 1 GeV protons of 4MW beam power. In Asia China is considering to build Beijing ISOL (formerly called Beijing CARIF) driven by an existing research reactor. Despite the difference in their drivers, the performance of Beijing ISOL is expected to be comparable to EURISOL, i.e., ~100 times more intense than the soon-to-be-ready ISOL facilities mentioned. Although there is no regional consensus in Asia, Beijing ISOL can be called AsianiSOL (ASOL).

RI beam with projectile fragmentation: Table 3 also includes the RI-beam facilities with projectile fragmentation. Compared to the ISOL facilities the projectile fragmentation facilities are superior in producing a wide range of rare RI beams, free from their chemical characteristics and short lifetime. Among these facilities, RIBF at RIKEN currently delivers the most powerful RI beams, typically $3*10^6$ of $^{132}$Sn per second with energy of 200 MeV/nucleon.

FRIB at MSU is truly the next generation projectile-fragmentation facility with 200 MeV/A 400kW LINAC followed by a fragmentation separator. FRIB will be completed by 2022 (partial start slated for 2020). Future options are to double the energy and to add an ISOL facility. Koreans have come up with the Radioactive Isotope Science Project (RISP) project to be launched in 2020. RISP and FRIB are aiming for a similar level of performance, $10^8$/s of $^{132}$Sn.

FAIR at GSI is another large facility. Using SIS100 with a U28+ 1.5GeV/A beam, it can improve the RI beam intensity by a factor of 100 to 1000 compared to the present SIS-based facility, making the facility comparable or superior to RIKEN RIBF.

Until Beijing ISOL becomes available China operates two facilities: BRIF2 (Beijing RI Facility) and Heavy Ion Research Facility in Lanzhou (HIRFL), and is starting to build Heavy Ion Accelerator Facility (HIAF) not in Lanzhou but in Guangdong Province (100km north-east of Hong Kong).

Superheavy element search : On December 30, 2015, IUPAC announced the discovery of the new elements 113, 115, 117, and 118. On November 8, 2016, the 7$^{th}$ row of the periodic table was completed with Nihonium, Moscovium, Tennessine and Oganesson. The future race hunting for the 119$^{th}$ and 120$^{th}$ elements will continue at the SHE factory at Dubna (to be completed in 2018) and RIKEN (introducing superconducting cavities to its linear accelerator by 2018). It should be noted that the future race requires actinide targets which can only be delivered by the High Flux Reactor at ORNL.

*****

In summary, one recognizes from Table 2 that the "Hot QCD" and "Cold QCD" facilities are shared efficiently worldwide. Compared to twenty years ago, there are fewer "Cold QCD" facilities because they have been left for collider projects. By having an electron-ion collider in the USA in the future, scientific coverage will be drastically expanded in parallel with more well-balanced regional interests and responsibilities. Assisted by the international competition among the rival facilities in the European, American and Asian continents, RI-beam facilities have become prevalent. Huge advances in this field of physics are expected in the coming 10-20 years. One may need to consider international amalgamation of research interests when either EURISOL or Beijing ISOL (formerly known as CARIF), both multi-billion dollar projects, is eventually realized.

| Type | Facility | Beam | | Target(ISOL) or Beam current(PF) | | Post acceleration | | EXP Start |
|---|---|---|---|---|---|---|---|---|
| | | Beam | Beam Power (kw) | Direct/ Conv/ PF | Fissions/s Beam pnA | MeV/A | $^{132}$Sn/s | |
| ISOL Coming | ARIEL | e 50MeV 10000mA p 500MeV 100mA | ~100 | Direct | $1*10^{14}$ | | $2*10^9$ | 2018 |
| | HIE ISOLDE | p 1GeV 2mA | 2 | D&C | $4*10^{12}$ | 5-10 | $2*10^8$ | 2017 |
| | SPIRAL2 | d 40MeV 5000mA | 200 | Conv | $1*10^{14}$ | 3-10 | $2*10^9$ | 2018 |
| | SPES | p 40MeV 200mA | 8 | Direct | $1*10^{13}$ | 10 | $3*10^8$ | 2021 |
| Super ISOL | EUR ISOL | p 1GeV 5000 mA | 4M | D&C | $1*10^{15}$ | 20-150 | $4*10^{11}$ | ? |
| | Beijing ISOL | Reactor | 6M | reactor | $2*10^{15}$ | >100 | $5*10^{10}$ | ? |
| PF Coming | FRIB | U+33 200MeV | 400 | PF | 8300 pnA | - | $10^8$~$10^9$ | 2020 |
| | RISP | U+79 200MeV | 400 | PF | 8000 pnA | - | $10^8$~$10^9$ | 2020 |
| | FAIR | U+28 1500MeV | 10 | PF | 50 pnA | - | $10^7$~$10^8$ | 2025 |
| PF Running | RIBF 2015 | U+86 345MeV | 4 | PF | 100 pnA | - | $3*10^6$ | running |

# Nuclear Power

**Updated February 28, 2018, by Nicolas Alamanos and Sylvie Leray (IRFU, DPhN, CEA-Saclay)**

Although more slowly than before, the energy demand is continuously increasing due to the growth of the world population and of the standard of living in developing countries and despite the efforts for energy saving and improved energy efficiency [1]. The world primary energy consumption has increased in 2016 by 1.0% following a ten-year average of 1.8% per year. In the World Energy Outlook 2017 (WEO 2017) [2] the International Energy Agency (IEA) estimates that the global energy needs will expand by 30% between 2017 and 2040, with a population growing from 7.4 billion to more than 9 billion. Two-third of the energy demand growth will come from Asia, in particular China and India.

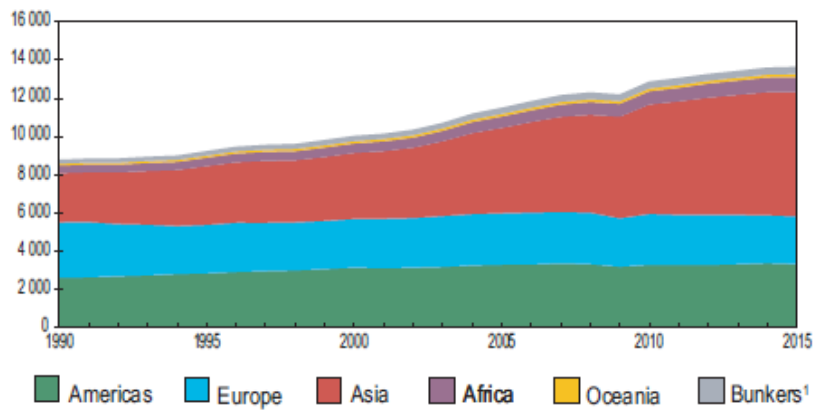

Figure 1 - World total primary energy supply by geographical region. From [3].

The need for electricity is rising even faster, 2.2% in 2016 following a ten-year average of 2.8% per year [1]. This is driven by the greater use of computers and smart electronic devices in developed countries and by more people getting access to electricity in developing countries. This trend will continue and even intensify with the developments of electric vehicles. In [2], IEA forecasts that electricity will represent 40% of the rise in final consumption to 2040. Again the demand growth will come mainly from Asia. It is for instance estimated that by 2040 the electricity needed for cooling in China will be larger than the total electricity supply of Japan in 2017.

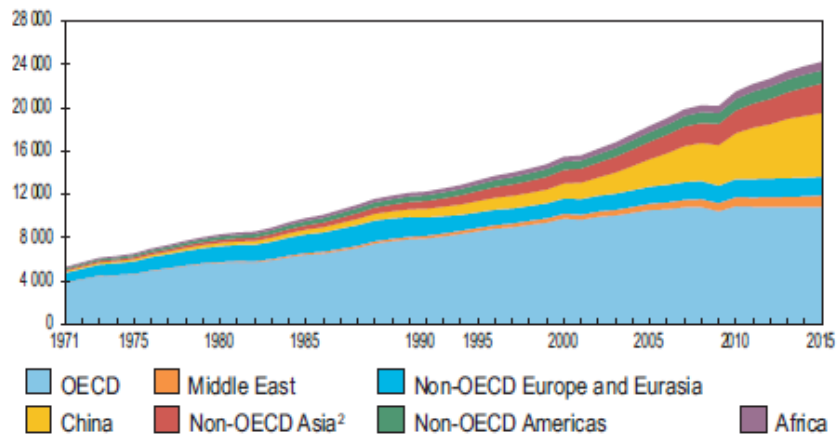

Figure 2 - World electricity generation from 1971 to 2015 by region (TWh). From [3].

At the same time, concern over the risk of climate change due to $CO_2$ emission is raising. The 2016 Paris Climate Agreement [4] aims at tackling the global climate change risk by limiting the global temperature rise below 2 degrees Celsius above pre-industrial levels. This implies reducing as far as possible the use of fossil fuels and therefore developing all possible alternative options such as energy saving, increasing energy efficiency and developing $CO_2$-free energy production. Because of the negative image of nuclear power, most of the countries are aiming at reducing their $CO_2$ emissions by increasing the share of renewable energies, i.e. mainly solar and wind energies since hydro-power deployment is limited by the availability of new sites. IEA foresees that the share of all renewables in total power generation will reach 40% in 2040, solar photovoltaics becoming the largest source of low-carbon capacity, driven mainly by China and India [2]. Fig. 3 (from [2]) shows that already in the period 2010-2016 new power plants were mostly renewables and that the tendency is expected to increase between 2017 and 2040.

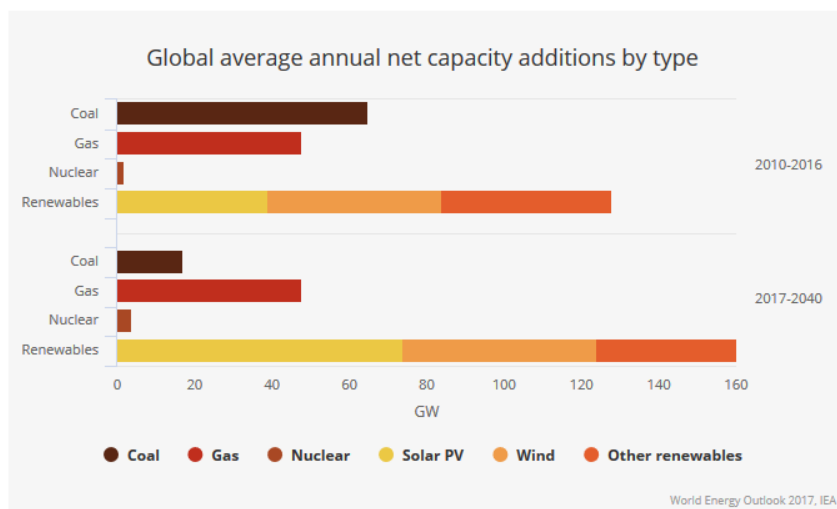

Figure 3 - Global average annual net capacity additions by type of fuel during the 2010-2016 period and foreseen by IEA between 2017 and 2040. From [2].

However, solar and wind energies have serious drawbacks, the main one being their variability and intermittency in the absence of effective storage solutions. This limits the share of power from renewable source in a grid and implies back-up solutions from other, often fossil fuels, sources and new smart and efficient grid management. Price, which is still much higher than for fossil fuels, is constantly decreasing and may not be an impediment to a rapid growth. A more serious concern is that solar photovoltaics and other renewable technologies are highly dependent on rare earth elements, which carry a risk of possible future supply disruption.

However, since 2013 the tendency has been inverted and nuclear power production is slowly but continuously increasing, driven mainly by China and India. Global nuclear power generation increased by 1.3% in 2016 with China accounting for all of the net growth. Presently, around 11% of the world's electricity is generated by about 450 nuclear power reactors. 58 reactors are under construction, equivalent to 16% of existing capacity, while an additional 150-160 are planned, equivalent to nearly half of existing capacity [5]. 20 of the 58 reactors in construction are in China, 6 in India and in Russia and the number of planned reactors is respectively 39 in China, 19 in India and 26 in Russia [5]. In addition, several new countries are moving towards nuclear energy, building their first reactors like Bangladesh, Belarus and the United Arab Emirates. WEO 2017 estimates that nuclear electricity generation will double by 2040.

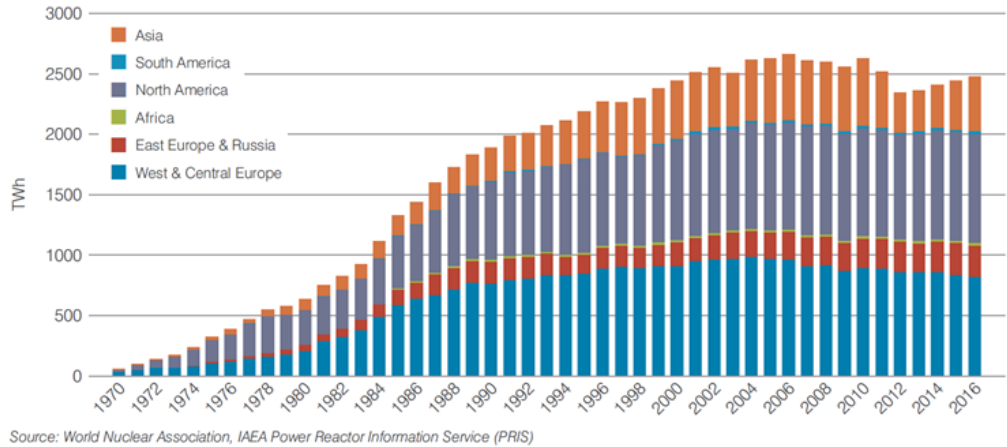

Figure 4 - Nuclear energy production from 1970 to 2016 by region (TWh). From [5]

Most of the reactors in operation are based on technology developed during the 1950s and later improved (Generation II). They are predominantly water-cooled reactors, either pressurized water (PWR) or boiling water reactors (BWR). The so-called Generation III reactors, which are the reactors presently being built, are using the same technology but their design has been optimized in order to reduce their cost by increasing the availability and lifetime, use more efficiently the fuel by allowing higher burn-up, and have an improved inherent safety.

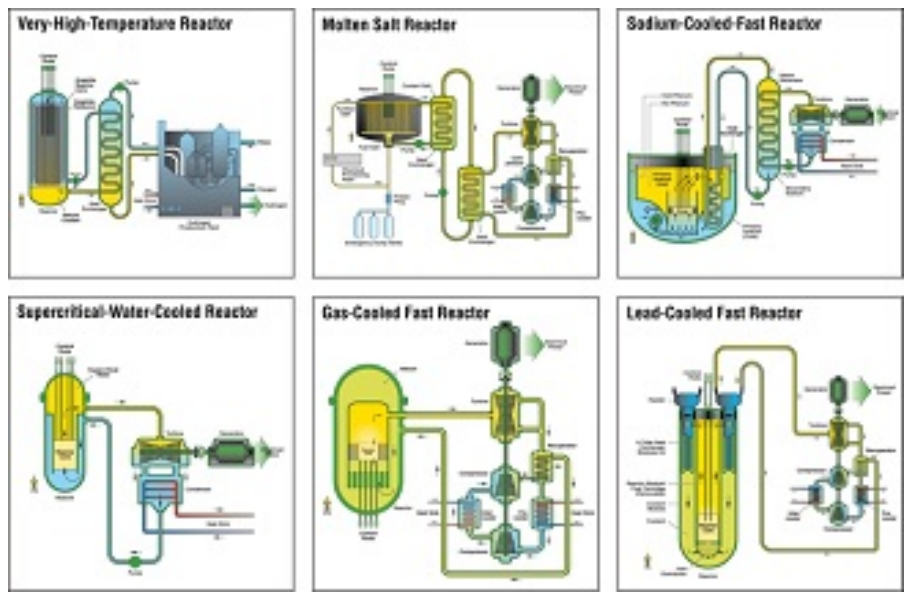

Figure 4 - The six technologies investigated in GIF. From [6]

A new generation of more innovative reactors is being investigated in the framework of the Generation IV International Forum (GIF) composed of representatives from 14 countries. Technology goals have been defined: Generation IV reactors should provide sustainable energy generation and long-term availability of systems, minimize and manage their nuclear waste, be economically competitive, have high level of safety and reliability, and be proliferation-resistant. Six different systems are studied. They use different coolant, light water for one system, lead-bismuth, sodium or fluoride salt and helium for two of them. Four are fast neutron reactors. The 2014 GIF Technology Roadmap Update has focused on the most relevant developments for a deployment in the next decade, which are the sodium-cooled fast reactor, the lead-cooled fast reactor and the very high temperature reactor technologies (VHTR) [6]. China is developing a prototype of VHTR, France and Russia are working on advanced sodium-fast reactor designs and a prototype of a lead fast reactor is also expected to be built in Russia.

Like any industrial activity, nuclear energy generation produces wastes but the management of highly radioactive nuclear wastes is a subject of great concern for the public and in fact one of the main reasons, with the fear of accident, why nuclear energy is not well accepted. Although in any case a final deep geologic disposal of remaining long-lived high level wastes will be necessary, the strategy regarding the management of spent fuel varies from one country to another [7]. Some countries, like US, Sweden or Finland, are planning to store directly the spent fuel while other ones, like France, Japan, Russia or China, reprocess the spent fuel to extract plutonium and make MOX fuels that are re-used and the remaining high-level wastes, composed mainly of fission products and minor actinides, are vitrified and intended to be sent to the geological disposal. While reprocessing reduces the amount, volume and radiotoxicity of the high-level waste (HLW) packages to be stored, it generates additional volumes of intermediate wastes during the reprocessing and fuel fabrication processes. As regards low (LLW) and intermediate-level (ILW) wastes, near-surface repositories are generally considered and already implemented for in several countries, including Czech Republic, Finland, France, Japan, Netherlands, Spain, Sweden, UK, and USA for LLW and Finland and Sweden for LLW and short-lived ILW. For HLW deep geological sites, the situation varies from, such as Sweden and Finland that have already selected a site and began to build the repository, to countries still in the selection process, for instance, UK and Canada. In France and US, a site has been chosen but the construction not yet started.

In parallel to the investigation of disposal solutions, extensive research is being conducted on partitioning and transmutation (P&T) that could lead to reducing the radiotoxicity and volume of the wastes to be finally stored. The idea is to change long-lived isotopes into stable or short-lived ones that could be stored for a limited period.

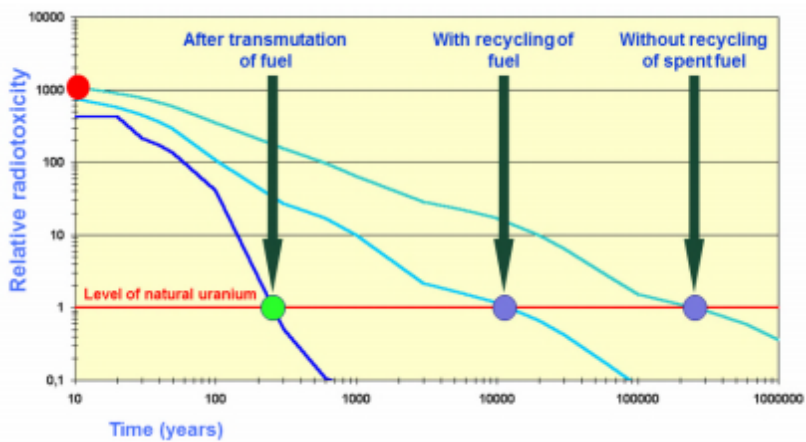

Figure 5. - Relative radiotoxicity on nuclear waste compared to natural uranium ore as a function of time after unloading from the reactor, for spent fuel without recycling, with Pu recycling and after transmutation of minor actinides. From [8, 9].

It is generally admitted that the transmutation of long-lived fission products is not viable due to the cost and the necessity in most case of isotope separation to avoid creating new undesirable isotopes while transmuting other ones. In contrast, transmuting minor actinides (MA) in addition to plutonium, lead to a significant reduction of the long-term radiotoxicity, as shown in Fig. 6. In Europe, in particular, the European Commission has been funding many projects concerning new types of reactors, fuels, and material involving minor actinides. A review of the state-of-the-art can be found in the OECD Nuclear Energy Agency (NEA) series of biennial Information Exchange Meetings on Actinide and Fission Product Partitioning and Transmutation [10]. Two different strategies for the transmutation of MA are envisaged: either adding a small amount of MA in a large number of commercial reactors or building a small number of dedicated units able to burn a large amount of MA. Indeed, the addition of MA into fuels degrades the safety parameters, such as the delayed neutron fraction or the Doppler coefficient, and therefore the amount that can be incorporated in a classical reactor is limited but could be counterbalanced by the number of reactors. Actually, studies of Generation IV reactors also include the investigation of the possibility to burn MA. Transmuting a large amount of MA may be possible in accelerator-driven systems (ADS) in which the operation in a sub-critical mode allows loosening on the safety parameters. There are currently three projects of ADS under studies: MYRRHA [11] in Belgium, C-ADS [12] in China and one in India, the latter being based on thorium fuel.

The development of new types of reactors, as generation IV or for the transmutation of MA, implies new types of coolant, moderator (if any), fuel and structure materials, as well as possibly different neutron flux environment. This, together with the requirement for reducing uncertainties, means that there is a need for the measurement of new or more precise nuclear data.